\documentclass{aa}  
\usepackage{graphicx}
\usepackage[bookmarks=open,colorlinks=true,linkcolor=blue,urlcolor=blue]{hyperref}
\usepackage{txfonts}
\usepackage{natbib}
\bibpunct{(}{)}{;}{a}{}{,} 

\newcommand{\xmm}{{\sl XMM-Newton}}
\newcommand{\cxo}{{\sl Chandra}}
\newcommand{\ex}{{EX~Lup}}
\newcommand{\vor}{{V1647~Ori}}
\newcommand{\vonze}{{V1118~Ori}}

\begin{document}

\title{A few days before the end of the 2008 extreme outburst of
EX~Lup :\\
accretion shocks and a smothered stellar corona\\ 
unveiled by XMM-Newton}
\titlerunning{accretion shocks and a smothered stellar corona unveiled by XMM-Newton}


   \author{N.\ Grosso\inst{1}
\and K.\ Hamaguchi\inst{2}
\and J.~H.\ Kastner\inst{3}
\and M.\ Richmond\inst{3}
\and D.~A.\ Weintraub\inst{4}
}
        
   \offprints{N.\ Grosso}
        
   \institute{Observatoire Astronomique de Strasbourg, Universit{\'e} de
Strasbourg, CNRS, UMR 7550, 11 rue de l'Universit{\'e}, F-67000 Strasbourg, France\\
              \email{\tt nicolas.grosso@astro.unistra.fr}
         \and NASA/Goddard Space Flight Center, Greenbelt, Maryland 20771, USA
         \and Rochester Institute of Technology, Rochester, New York 14623-5604, USA
	 \and Vanderbilt University, Nashville, Tennessee 37235, USA
            }

   \date{Received 10 December 2009 / Accepted 21 June 2010}


  \abstract
   {\ex~is a pre-main sequence star that exhibits repetitive and
 irregular optical outbursts driven by an increase in the  mass
 accretion rate in its circumstellar disk. In mid-January 2008, \ex, the prototype
 of the small class of eruptive variables called EXors, began an
 extreme outburst that lasted seven months.}
   {We attempt to characterize the X-ray and UV emission of \ex~during this
outburst.}
   {We observed \ex~during about 21~h with \xmm, simultaneously in
X-rays and UV, on August 10--11, 2008 -- a few days before the end of its
2008 outburst -- when the optical flux of \ex~remained about 4 times
above its pre-outburst level.}
   {We detected \ex~in X-rays with an observed flux in the 0.2--10~keV energy range of
$5.4\times10^{-14}$~erg\,s$^{-1}$\,cm$^{-2}$ during a low-level
period. This observed flux increased by a factor of four during a
flaring period that lasted about 2~h. The observed spectrum of the
low-level period is dominated below $\sim$1.5~keV by emission from a
relatively cool plasma ($\sim$4.7~MK) that is lightly absorbed 
($N_\mathrm{H}\simeq3.6\times10^{20}$~cm$^{-2}$) and above
$\sim$1.5~keV by emission from a plasma that is $\sim$ten times hotter
and affected by a photoelectric absorption that is 75 times larger. The
intrinsic X-ray luminosity of the relatively cool plasma is
$\sim4\times10^{28}$~erg\,s$^{-1}$. The intrinsic X-ray luminosity
of \ex, $\sim3.4\times10^{29}$~erg\,s$^{-1}$, is hence
dominated by emission from the hot plasma. During the X-ray
flare, the emission measure and the intrinsic X-ray luminosity of this
absorbed plasma component is five times higher than during the
low-level period. We detected UV variability on timescales
ranging from less than one hour up to about four hours. We show from
simulated light curves that the power spectral density of the UV light
curve can be modeled with a red-noise spectrum with a power-law index
of $1.39\pm0.06$. None of the UV events observed on
August 10--11, 2008 correlate unambiguously with simultaneous X-ray peaks.}
   {The soft X-ray spectral component is most likely associated
with accretion shocks, as opposed to jet activity, given the absence of
forbidden emission lines of low-excitation species (e.g.,
[\ion{O}{i}]) in optical spectra of \ex~obtained during outburst.
The hard X-ray spectral component, meanwhile, is most likely associated with a
smothered stellar corona. The UV
emission is reminiscent of accretion events, such
as those already observed with the Optical/UV Monitor from other
accreting pre-main sequence stars, and is evidently dominated by emission from
accretion hot spots. The large photoelectric
absorption of the active stellar corona is most likely due to high-density gas
above the corona in accretion funnel flows.}

   \maketitle
%
\section{Introduction}

Young, low-mass stars that accrete material through their
circumstellar disks can exhibit dramatic increases in brightness in
the optical. These erupting pre-main sequence (PMS) stars are
classified as FUors and EXors, named after the prototypes
FU~Ori and \ex. FUors exhibit a single (observed) major
increase in brightness ($\ge$4--5\,mag), occurring perhaps once per
century or only a few times over the PMS lifetime of the star,
with a decay timescale of 10--100\,yr \citep{herbig66,herbig77,hartmann96}.
The characteristics of the EXor class are
dominated by what is known about the prototype, namely, that the
outbursts of \ex~are repetitive and irregular, occurring perhaps every
few decades, with a decay timescale of a few months to a few years
\citep{herbig89,herbig08}. In PMS stars, the typical rates of
mass accretion are lower than
$\sim$$10^{-7}$\,$M_\odot$\,yr$^{-1}$, but, because of thermal
instability, may jump to
$\sim$$10^{-6}$--$\sim$$10^{-4}$~$M_\odot$\,yr$^{-1}$, producing an
FUor-type or EXor-type optical outburst \citep{hartmann96}.

Although luminous X-ray emission is characteristic of PMS stars
(see reviews of \citealt{feigelson99} and \citealt{guedel09}), there
exist very few observations of high-energy (UV through  X-ray)
radiation from FUors or EXors in outburst. 
An \xmm~observation of FU Ori itself shows the cool ($\sim 8$~MK) and
hot ($\ge 58$~MK) plasma components usually detected in T~Tauri stars
\citep{preibisch05b}, but the absorption toward the hot plasma
component is at least 10 times larger than the optical extinction
\citep{skinner06}. The observed X-ray flux is dominated by the hot
plasma component, which is evidently due to magnetic
activity. However, \cite{skinner06} contend that the temperature of
the cool component is too high to be caused by accretion shocks.
An \xmm~observation of the FUor V1735~Cyg detected a hard
X-ray spectrum from a hot ($\ge 58$~MK) plasma
\citep{skinner09}. In these two FUors, the high temperature of the hot
plasma component indicates that the X-ray emission is dominated by
magnetic processes. Pre-outburst observations of these FUors in X-rays
are not available, hence we are unable to assess the impact of the accretion
outburst on their X-ray emission.

In late 2003, \vor, a young, low-mass star deeply embedded in the
L1630 cloud in Orion ($d\sim400$~pc) brightened suddenly, illuminating
McNeil's nebula. 
\vor~is the first PMS star to have been
observed in X-rays, before, during, and after a major mass-accretion
episode \citep{kastner04,kastner06,grosso05,grosso06c}. \cxo~and
\xmm~observations demonstrate that the sharp increase in X-ray flux
post-outburst relative to pre-outburst closely tracked that of the
optical/IR brightness increase of \vor. The X-ray mean flux of
\vor~follows the optical/IR decline of the outburst, but enhanced
variability on timescales of less than a day was also observed in X-rays. 
These results appear to be most accurately explained by star-disk
magnetic reconnection events that
were generated in association with this major mass-accretion episode.
\vor~returned to near pre-outburst
levels only by late 2005, and was reported to flare again in the
optical in late August, 2008 \citep{itagaki08}. A new snapshot
monitoring with \cxo~(Cycle~10; D.~A.\ Weintraub, PI) caught the
overall rise/fall shape of this new X-ray outburst (Teets et al.\
2010a, in preparation). A Suzaku observation in October 2008 detected
strong fluorescent iron line emission that was not present during the
2003--2005 outburst. The structure of the
circumstellar gas very close to the stellar core that absorbs and
re-emits X-ray emission from the central object may have changed
between 2005 and 2008 \citep{hamaguchi10}.

The EXor candidate \vonze, a young low-mass star in the
Orion Nebula ($d\sim400$~pc), displayed a mass accretion outburst from
2005 to 2006. Follow-up observations of this object detected a
moderate enhancement in the X-ray flux that was correlated with the
optical/IR flux \citep{audard05b,audard10,lorenzetti06}. 

The Herbig Be/FUor binary star Z~CMa exhibited in 2008--2009 the
largest optical outburst reported during the almost 90 years of available
observations. The FUor component is responsible for this optical
outburst and the X-ray emission of Z~CMa. However, no changes in the
stellar X-ray properties of the FUor component were seen during this
optical outburst \citep{stelzer09}. \cite{szeifert10} show that the
bolometric luminosity of Z~CMa remained surprisingly constant during
this optical outburst and conclude that the increase in luminosity
was caused by a decrease in extinction.

In contrast to the previous results, 
a \cxo~observation was unable to
detect the outburst source that illuminates the new nebula associated with
IRAS~04376+5413 in LDN~1415, located at $d \sim
170$~pc \citep{kastner06b,stecklum06,stecklum07}.

Given the variety of behavior observed during PMS star outbursts,
additional monitoring of PMS star outbursts is essential. We are conducting such a monitoring campaign of potential FUor
and/or EXor outbursts, by means of coordinated \xmm~and \cxo~target of
opportunity (ToO) programs. Here, we report on UV/X-ray observations
of \ex~obtained via our \xmm~ToO program.

\ex~is located between the star-forming clouds
Lupus~3 and 4 \citep[see review on the Lupus clouds of
][]{comeron08}, at a distance of $155\pm8$~pc \citep{lombardi08}. It
has a normal brightness of $V=13.2$~mag \citep{herbig07} and 
a M0V spectral type \citep{herbig77b}. In 1955, the
visual observations of A.\ F.\ Jones helped to infer that \ex~had
experienced a major outburst with a peak magnitude
of $V=8.4$~mag \citep{herbig77}. During the 1993--2005 period, eight outbursts
not brighter than $V=11$~mag were observed
\citep{herbig01,herbig07}. \cite{jones08} announced that \ex~was again
in outburst with onset in mid-January 2008.
Follow-up visual observations showed that in early
February~2008 \ex~reached a peak $V$-band luminosity of $\sim$8~mag,
making this last outburst the brightest ever observed from
\ex~\citep{jones08b,kospal08}. On the basis of {\sl Spitzer}
spectroscopy obtained before and during outburst, \cite{abraham09}
claim that heat from this bright outburst transformed, via thermal annealing,
amorphous silicate dust in the surface layer of the inner accretion
disk of \ex~into crystalline form (namely, forsterite). They propose that
interstellar (amorphous) dust within the protosolar nebula was transformed
into crystalline silicate grains present in comets and meteorites
during these accretion outbursts.

The 2008 optical outburst
of the EXor prototype \ex~triggered our (anticipated) ToO observations
with \cxo~(Cycle~9; D.~A.\ Weintraub, PI). This \cxo~monitoring of
\ex~with snapshots showed that \ex~was
relatively X-ray bright in March 25, 2008, and had then faded by a factor
of three by June 16, 2008, and by another factor of about two by October
2008. The changes in X-ray brightness appear to correlate
in a meaningful way with the optical/IR fluxes, which strongly
suggests that the level of X-ray production above quiescence for this
star is generated by the same physical process that produced the
optical eruption, i.e., mass accretion (Teets et al.\ 2010b, in preparation).
The positive detection of \ex~in X-rays with
\cxo~allowed us to trigger a long-exposure observation with \xmm~(AO-7; N.\
Grosso, PI) to perform X-ray CCD spectroscopy and to investigate X-ray
and UV variability on a timescale of about one day that we report
here. We describe in Sect.~\ref{obs} the \xmm~observation and our
reduction of these data. We present our results in
Sect.~\ref{results}, discuss them further in Sect.~\ref{discussion},
and offer our conclusions in Sect.~\ref{conclusions}. We describe in
detail our reduction of the Optical/UV Monitor data in
Appendix~\ref{appendix:drifts}.

\section{XMM-Newton observation and data reduction}
\label{obs}

\subsection{Observing period}

\begin{figure}[!t]
\centering
\includegraphics[height=\columnwidth,angle=90]{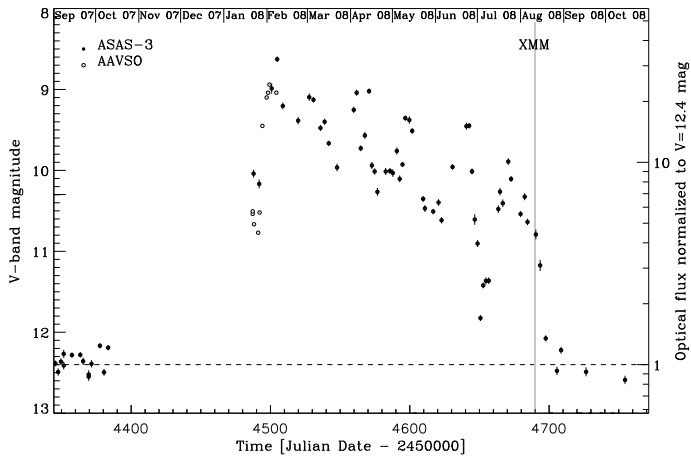}
\caption{$V$-band light curve of \ex. Filled and empty circles are
$V$-band magnitudes of \ex~from the All Sky Automated Survey
(ASAS-3) and the American Association of Variable Star Observers
(AAVSO), respectively. The horizontal dashed line is the
pre-outburst optical level. The right y-axis gives the optical flux in
pre-outburst flux units. The grey stripe at $JD\simeq2454690$ identifies
the \xmm~observing period in August 2008.}
\label{fig:lc_V}
\end{figure}

The 78~ks exposure was scheduled
at the very beginning of the Summer visibility window of
\ex~(\xmm~revolution 1588), on August 10--11, 2008. This observing
period is shown in the context of the 2008~extreme outburst of \ex~in
Fig.~\ref{fig:lc_V}. 

For the $V$-band data plotted in
Fig.~\ref{fig:lc_V}, we selected only measurements with error bars
from the American Association of Variable Star Observers
(AAVSO)\footnote{See the AAVSO site at
\href{http://www.aavso.org/}{\tt http://www.aavso.org}\,.} and from
the ASAS-3 photometric catalog\footnote{The ASAS-3 photometric
catalog is available at \href{http://www.astrouw.edu.pl/asas/}{\tt
http://www.astrouw.edu.pl/asas/}\,.} of the All Sky Automated Survey
\citep[ASAS;][]{pojmanski02}. We estimated
the pre-outburst optical level of \ex~using the median of $V$
measurements obtained on Sep.--Oct.\ 2007, just before the conjunction
of \ex~with the Sun. We found that $V\simeq12.4$~mag, which is above its
normal brightness of $V=13.2$~mag \citep{herbig07}. \ex~returned to
this pre-outburst optical level on late
August, 2008.

Figure~\ref{fig:lc_V} shows that our \xmm~observation was obtained only
a few days before the end of the 2008 outburst, but that \ex~was
clearly still in an elevated state during the observation. 
Ten hours after the end of the \xmm~observation,
\ex~was observed by the ASAS at $V=10.79\pm0.06$~mag. For
comparison, this optical level is as bright as the peak of the brightest
outburst of the 1993--2005 period that was observed on August, 2002, and 4.4
times above its 2008 pre-outburst optical level.

\subsection{Instrumental setup}

The prime instrument was EPIC, with pn
\citep{strueder01} and the two MOS \citep{turner01}, used in the full
frame science mode with the medium optical blocking filter. 

The Optical/UV Monitor \citep[OM;][]{mason01} was used in the default fast
mode with the UVM2 filter (covering the wavelength band from 205 to
245~nm, with an effective wavelength of 231~nm), which allowed us to
image a $17\arcmin\times17\arcmin$ field-of-view and obtain for
\ex~fifty UV event lists (1200~s observations separated by gaps of
321.6~s due to overhead time).
Because, during our OM observation, 
no good guide stars were found, we were unable to use the \xmm~Science
Analyzing System (SAS)\footnote{See 
\href{http://xmm.esac.esa.int/external/xmm_data_analysis/}{http://xmm.esac.esa.int/external/xmm\_data\_analysis/}\,.}
to correct the UV event lists of \ex~for spacecraft drifts. We present in
Appendix~\ref{appendix:drifts} the method we used to correct the OM
fast-mode light curve for spurious features caused by spacecraft drifts.

\subsection{Target identification}

The pipeline data products (PPS) provide correlations of EPIC
sources in our observation with archival catalogs. The EPIC source
XMMU~160305.5-401825, with a positional uncertainties of $1\farcs56$
(including a $1\farcs50$ systematic error), is located only
$0\farcs7$ from the coordinate position of 2MASS~J16030548-4018254
(=\ex). Therefore, this EPIC source is the X-ray counterpart of
\ex. The closest neighboring X-ray source detected by \xmm~is located
at a distance of $2\farcm3$ from the \ex~source. The Cycle 9
\cxo~images show that the closest X-ray source to \ex~is a faint
object located at a distance of $44\arcsec$. Therefore, the X-ray and
UV flux from XMMU~160305.5-401825 are not contaminated by flux from
any nearby source.

From the PPS list of OM sources detected in the
$17\arcmin\times17\arcmin$ field-of-view, 
we correlated the positions of OM sources of a signal-to-noise ratio
(S/N) greater than 3.5 with the positions of 2MASS sources\footnote{We use the
SAS task {\tt eposcorr}.}, which corrects the OM source positions for a
small pointing offset ($\Delta\alpha=1\farcs6\pm0\farcs2$ and
$\Delta\delta=-2\farcs7\pm0\farcs2$), leading to a position
uncertainty of $0\farcs3$. We find a bright UV source located only $0\farcs3$ from
2MASS~J16030548-4018254. Therefore, this OM source is the UV
counterpart of \ex.

\subsection{Data reduction}

The data were reprocessed and analyzed using the SAS (version
9.0) following standard procedures\footnote{See SAS threads at
\color{blue}{\tt http://xmm.esac.esa.int/sas/current\-/documentation/threads/}\,.}.
Source plus background X-ray events\footnote{For PN, we selected only
single and double pixel events (i.e., {\tt PATTERN} in the 0 to 4
range) with {\tt FLAG} value equal to zero. For MOS, we selected
single, double, triple, and quadruple pixel events ({\tt PATTERN} in
the 0 to 12 range), and used the FLAG list {\tt \#XMMEA\_SM}, which is
advised for accurate spectral analysis (see the SAS task
\href{http://xmm.esa.int/sas/current/doc/emchain/node4.html}{emchain})}. were
extracted within a circular region centered on \ex~that had been optimized to
maximize the S/N (which gives a $14\arcsec$ radius
for pn). The background X-ray events located on the same CCD were
extracted using an annular region centered on \ex. We applied to the
X-ray light curve of each camera both relative corrections (dead time,
GTIs, exposure, background
counts) and absolute corrections (vignetting, bad pixels, chip gaps,
PSF, quantum efficiency) with the SAS task {\tt
epiclccorr}. Therefore, we report the count rates that would have been
detected if the X-ray events had been collected with an infinite
extraction radius. 

All the EPIC spectra presented here were binned to at least 15 counts per
spectral bins. Spectral modeling was performed with XSPEC \citep[version
12.5.1;][]{dorman01}.  We used X-ray spectra from optically-thin plasmas in
collisional ionization equilibrium, which include continuum and
emission line output from the Astrophysical Plasma Emission
Code\footnote{For more informations on APEC, see
\color{blue}{http://hea-www.harvard.edu\-/APEC}\,.}
({\tt vapec} model in {\tt XSPEC}). We adopted the abundance
pattern of the {\sl XMM-Newton Extended Survey of the Taurus molecular cloud}
\citep[XEST;][]{guedel07}, which we take to be the typical elemental abundances
of the coronae of young stars, as measured with grating X-ray
spectroscopy\footnote{The abundance pattern of the XEST
\citep{guedel07} is~: C=0.45, N=0.788, O=0.426, Ne=0.832, Mg=0.263,
Al=0.5, Si=0.309, S=0.417, Ar=0.55, Ca=0.195, Fe=0.195, and Ni=0.195
\citep[with respect to the solar photospheric abundances
of][]{anders89}.}. We combined these emission spectra with
photoelectric absorption using the {\tt XSPEC} model {\tt wabs},
which is based on the photo-ionization cross sections of
\cite{morrison83} and the solar abundances of \cite{anders82}.

\subsection{Effective exposure}

\begin{figure}[!t]
\centering
\includegraphics[trim=0 0 0 330,clip,width=\columnwidth]{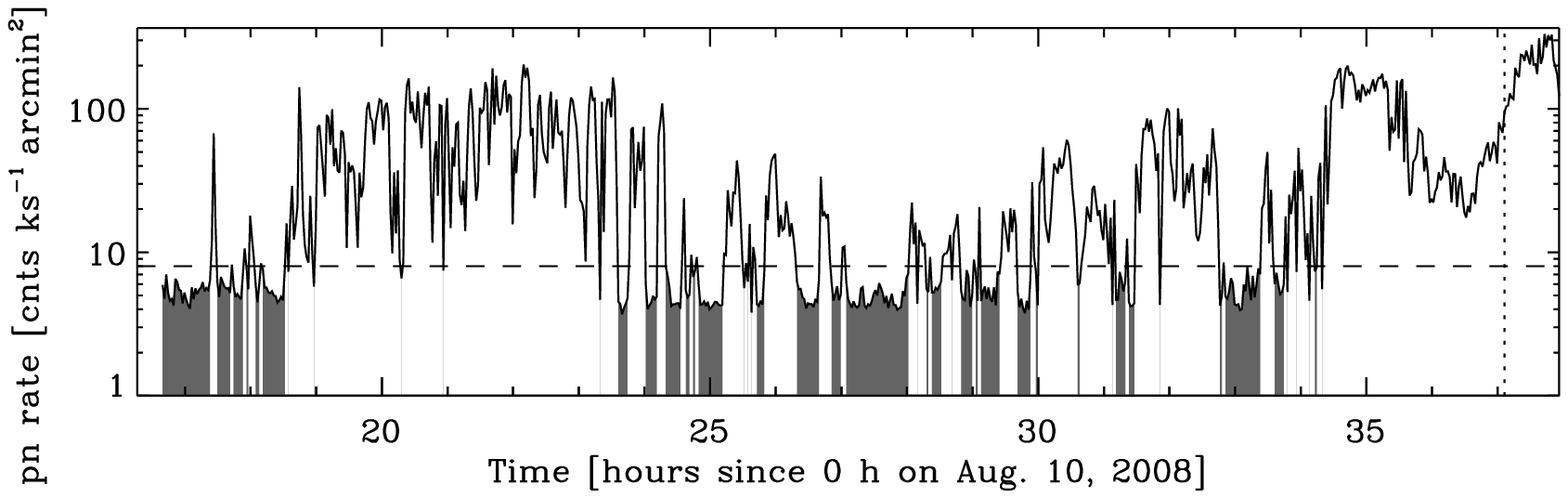}
\caption{Flaring-background light curve in the pn detector during our
\xmm~observation of \ex. The line is the background light curve
produced by the SAS task {\tt epchain}. The horizontal dashed line
indicates the maximum level of flaring-background for our spectral
analysis. The resulting time periods with low flaring-background are
shown in grey. The vertical dotted line indicates the end of the
useful exposure.}
\label{fig:lc_flaring_background}
\vspace{0.5cm}
%
%
\includegraphics[width=\columnwidth]{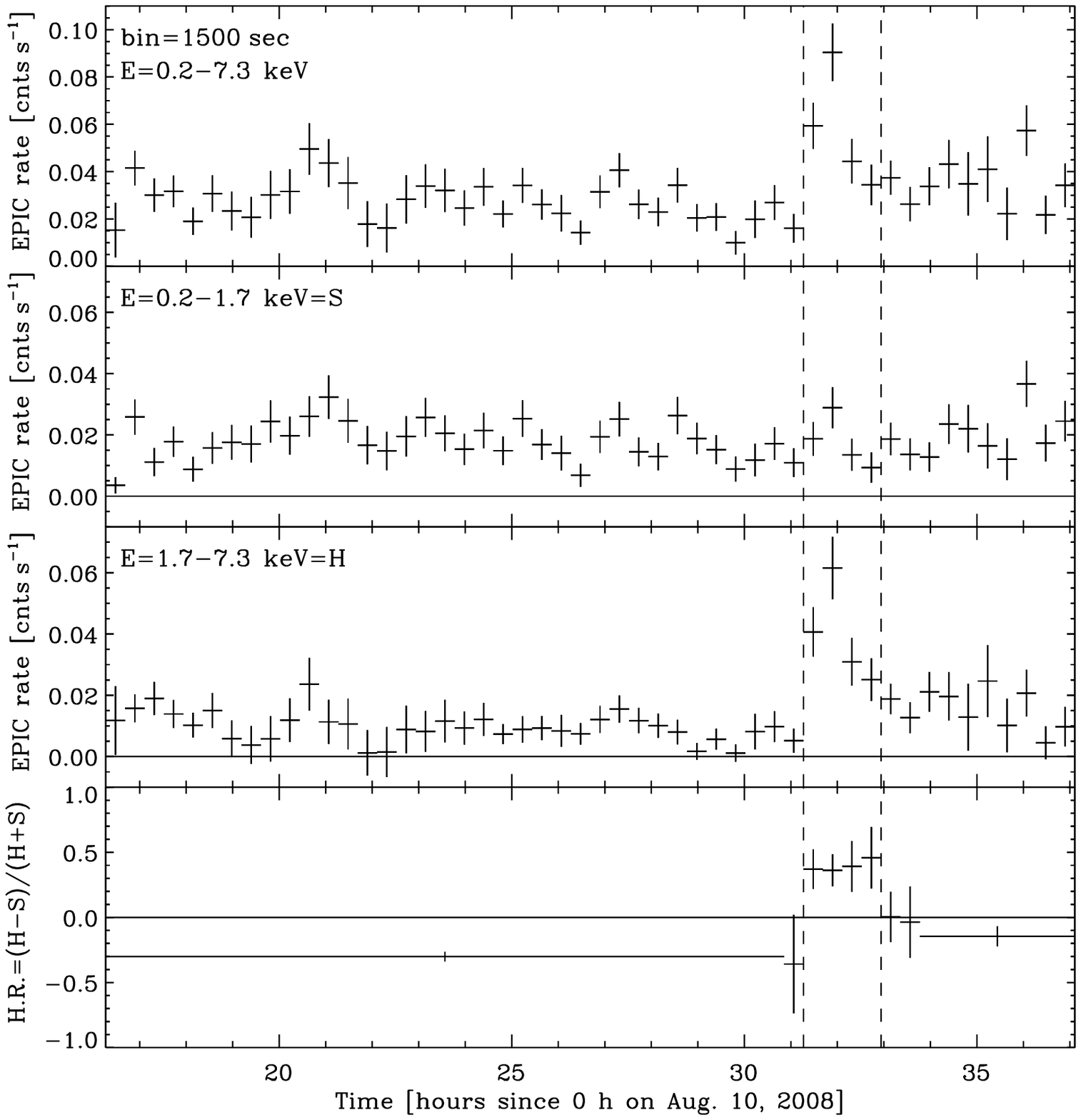}
\caption{\xmm/EPIC background-subtracted X-ray light curves
of \ex. {\sl The top panel} shows the EPIC (pn+MOS1+MOS2)
X-ray light curve of \ex~in the energy
band from 0.2 to 7.3~keV. {\sl The middle panels}
show the soft (S=0.2--1.7~keV) and hard (H=1.7--7.3~keV) band
X-ray light curves. The bin size of
the X-ray light curves is 1500~s. {\sl The bottom panel} shows the variation
of the corresponding hardness ratio. The vertical
dashed lines bracket the source flaring period.
}
\label{fig:lc_epic}
\end{figure}

This observation was affected by bad space
weather (Fig.~\ref{fig:lc_flaring_background}). In
particular, the last 3~ks of the exposure were strongly affected by
highly elevated flaring-background. Therefore, we did not use
EPIC data obtained on August 11, 2008 after about 13:00
(i.e., after about 37~h in Fig.~\ref{fig:lc_flaring_background}).

For the time variability study presented in
Sect.~\ref{x-ray_variability}, this \xmm~observation with an
effective exposure of $\sim74$~ks is the longest continuous X-ray
(and UV) observation that was obtained during this optical outburst of \ex.

For the spectral analysis presented in
Sect.~\ref{spectrum}, we suppressed time intervals with flaring
background rate above $8.0$~pn\,count\,ks$^{-1}$\,arcmin$^{-2}$,
which reduces the effective exposure for the pn spectrum to only
$\sim$22~ks (see time periods shown in grey in Fig.~\ref{fig:lc_flaring_background}). 

\section{Results}
\label{results}

\subsection{X-ray variability}
\label{x-ray_variability}

To limit the background, we concentrated our variability analysis on the
energy range from 0.2~keV to 7.3~keV. We divided this energy range into 
soft (S, 0.2--1.7~keV) and hard (H, 1.7--7.3~keV) bands to allow a 
hardness ratio study using the formula $HR=(H-S)/(H+S)$. We used 1500~s for
the bin size of the X-ray light curves of each camera to ensure that
each time bin has a positive count rate in at least one of the three
cameras. We replaced the negative count rates by zero.

Figure~\ref{fig:lc_epic} shows the background-subtracted
X-ray light curves of \ex~obtained by adding the pn, MOS1, and MOS2
light curves. During the first two-thirds of the observation, \ex~exhibited a
low level of activity, with amplitude variation of about
0.02~EPIC\,counts\,s$^{-1}$ around an average count rate of about
0.03~EPIC\,counts\,s$^{-1}$. Then, just after 31.2~h, as shown in
the top panel of Fig.~\ref{fig:lc_epic}, we observed a rapid increase in the
count rate that peaked at $\sim$3 times the average low-level count rate, 
before decaying to the pre-outburst level at the end of the
observation.
This last, decay phase, occurring just
after 33 h, was strongly affected by the highly elevated
background. The source flare-like event was detected mainly in
the hard band, and was the only clear variability detected in this energy
band. There was also a possible fainter second source flare-like
event seen in both the hard and soft X-ray bands around 20.5~h, but
this time interval was strongly affected by a large flaring background;
therefore, we cannot conclude definitively that \ex~experienced two
flares, though it certainly experienced only one bright flare peaking
above 0.07~count\,s$^{-1}$ during this 74~ks observing window.


\begin{table*}[!t]
\caption{Physical parameters of the X-ray emission from \ex~obtained
by spectral fitting of the pn data.}
\label{table:fit_epic_wabs_vapec_angr_xest_parameters}
\begin{tabular}{@{}cccccccccccccc@{}}
\hline
\hline
\noalign{\smallskip}

& \multicolumn{3}{c}{Soft spectral component} 
& \multicolumn{3}{c}{Hard spectral component}
\vspace{-0.2cm}\\

& \multicolumn{3}{c}{\hrulefill} 
& \multicolumn{3}{c}{\hrulefill} 
\\
$Period^{\,a}$
& $N_\mathrm{H,s}$
& \multicolumn{1}{c}{$T_\mathrm{s}$} 
& \multicolumn{1}{c}{$EM_\mathrm{s}$} 
& $N_\mathrm{H,h}$
& \multicolumn{1}{c}{$T_\mathrm{h}$} 
& \multicolumn{1}{c}{$EM_\mathrm{h}$} 
& \multicolumn{1}{c}{$F_\mathrm{X}^{\,b}$}
& \multicolumn{1}{c}{$L_\mathrm{X,s}^c$}
& \multicolumn{1}{c}{$L_\mathrm{X,h}^c$}
& \multicolumn{1}{c}{$\chi^2/\mathrm{d.o.f.}$} 
& \multicolumn{1}{c}{${\cal Q}^{\,d}$}
\\
                
& ($10^{20}$\,cm$^{-2}$)   
& \multicolumn{1}{c}{(MK)}  
& \multicolumn{1}{c}{($10^{51}$\,cm$^{-3}$)} 
& ($10^{22}$\,cm$^{-2}$)   
& \multicolumn{1}{c}{(MK)}  
& \multicolumn{1}{c}{($10^{51}$\,cm$^{-3}$)} 
& \multicolumn{1}{c}{(erg\,cm$^{-2}$\,s$^{-1}$)}
& \multicolumn{2}{c}{($10^{28}$\,erg\,s$^{-1}$)}
&
& 
\\
\noalign{\smallskip}
\hline
\noalign{\smallskip}
L & 3.6        & 4.7      & 4.2      & 2.7     & 53      & 18.3    & $0.54\times10^{-13}$ & 4.3 & 29.2 & 10.38/12 & 0.58\\
  & $\le$14.8  & 3.6--7.9 & 2.9--7.9 &0.4--7.3 & $\ge18$ & 7.9--114&\\
\noalign{\smallskip}
\hline
F & =3.6       & 4.2      & 5.7      & 5.2       & 61      & 102    & $2.3\times10^{-13}$ & 5.6 & 170 & 4.03/8 & 0.85 \\
  &            & 3.0--6.8 & 3.9--7.5 & 2.6--11.2 & $\ge19$ &
51--594&\\
\noalign{\smallskip}
\hline
\noalign{\smallskip}
\end{tabular}
\\
Notes:  The model used for the photoelectric absorption is {\tt
wabs}. The model used for the continuum and emission lines produced by
an optically thin plasma in thermal collisional ionization equilibrium
is {\tt vapec}, using the abundance pattern of the {\sl XMM-Newton
Extended Survey of the Taurus molecular cloud}
\citep[XEST;][]{guedel07}. The second line of each period
corresponds to the confidence intervals at the 90\% confidence level
(i.e., $\Delta\chi^2=2.706$ for each parameter of interest).
\begin{list}{}{}
\item[$^a$] The low-level and the flaring period are
labeled L and F, respectively.
\item[$^b$] Observed X-ray flux in the 0.2--10~keV energy
range.
\item[$^c$] X-ray luminosity corrected for the absorption in the 0.2--10~keV energy
range, at a distance of 155~pc \citep{lombardi08}.
\item[$^d$] The $\cal{Q}$-value is the probability that one
would observe the chi-square value, or a larger value, if the assumed
model is true, and the best-fit model parameters are the true parameter values.
\end{list}
\end{table*}

The bottom panel of Fig.~\ref{fig:lc_epic} shows that the
hardness ratio is soft during the low-level period, hard during the
main period of the source flare lasting about 6.5~ks (see
vertical dashed lines in Fig.~\ref{fig:lc_epic}), and returns to
softer values during the decay phase and the end of the
observation. We conclude that this variability and this hardening is
characteristic of an X-ray flare from \ex.

For the pn, MOS1, and MOS2 light curves in the soft
band, we applied a Kolmogorov-Smirnov (KS) test\footnote{We used
the {\tt ftools} task {\tt lcstats} ({\tt xronos} subpackage).
{\tt ftools} is available at 
\color{blue}{\tt
http://heasarc.nasa.gov/\-lheasoft/ftools/ftools\_menu.html}\,.}
to establish whether the variations are consistent
with Poisson noise that is associated with a constant
source. For each of the three light curves, we found  a KS-test
probability below 0.001, which indicates a
reliable level of soft-band variability \citep[e.g.,][]{getman05}.

\subsection{Time-dependent X-ray spectral analysis}
\label{spectrum}

\begin{figure}[!t]
\centering
\includegraphics[height=0.8\columnwidth,angle=90]{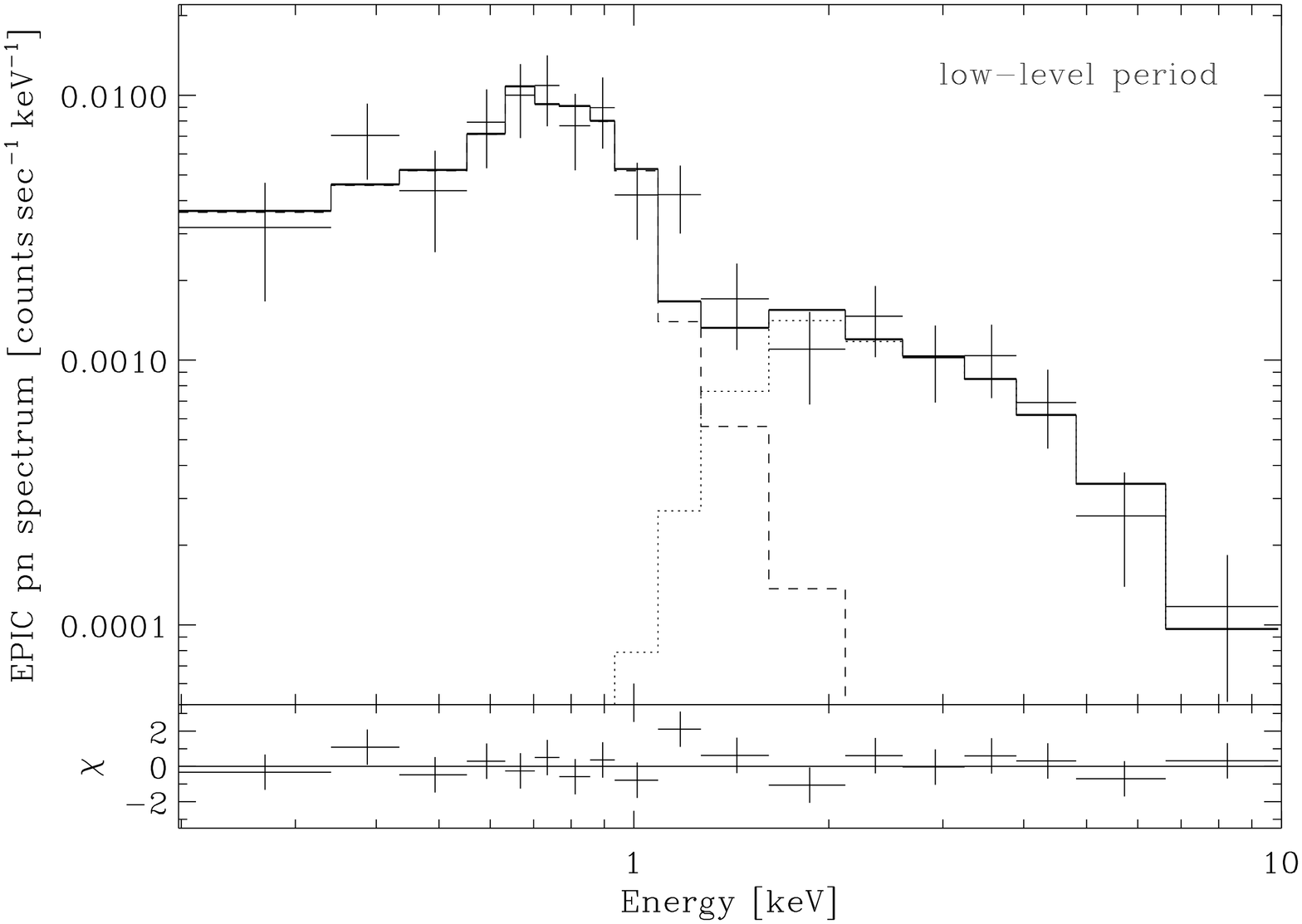}
\vspace{0.25cm}\\
\includegraphics[height=0.8\columnwidth,angle=90]{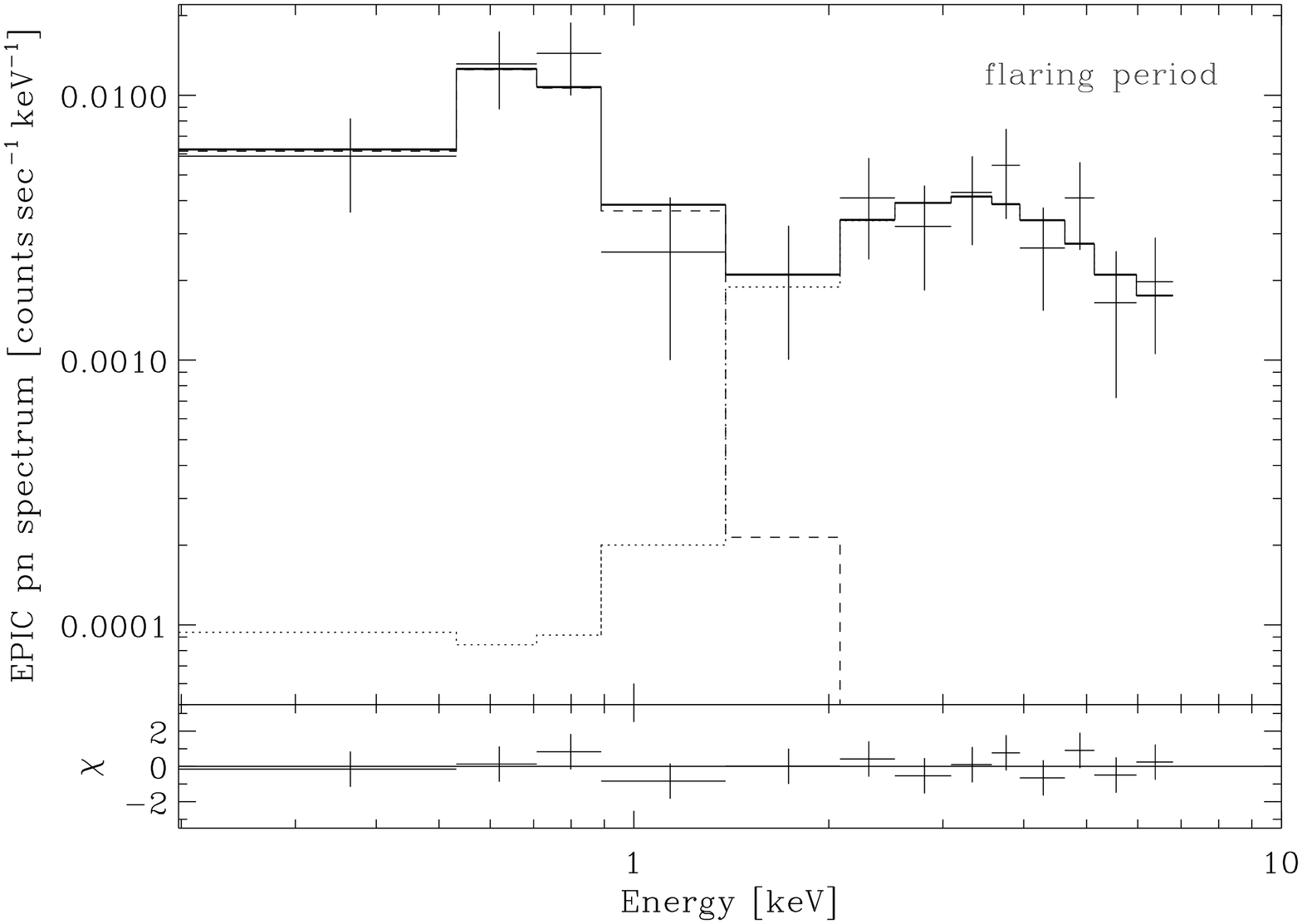}
\caption{pn spectra of \ex. {\sl Top panel:} low-level emission. {\sl
Bottom panel:} flaring emission. The continuous,
dashed and dotted lines are the best model of the whole spectrum, and
the soft and hard components (see
Table~\ref{table:fit_epic_wabs_vapec_angr_xest_parameters}),
respectively. The residuals of the fit (in sigma units) are shown in the
bottom sections of each panel.}
\label{fig:spectra}
\end{figure}

The MOS spectra of EX Lup are very faint hence cannot be used to
provide useful constraints from spectral modeling; the
spectral modeling results we present were therefore obtained
exclusively from the pn spectra.  
We first performed X-ray modeling of the data from the low-level
period of our observation, before the source flare that began at
31.2~h (see vertical dashed lines in Fig.~\ref{fig:lc_epic});
this low-level period has a low flaring-background exposure time of
18.6~ks (see Fig.~\ref{fig:lc_flaring_background}).
The top panel of Fig.~\ref{fig:spectra} shows the spectrum of the
low-level period. It is dominated by soft emission that peaks around
~0.7 keV, which suggests that the emission below $\sim$1.5~keV
originates in a relatively cool plasma. X-rays were detected at energies as low as
0.2~keV, which indicates that the photoelectric absorption is
low, for energies as low as 10~keV.

A single-component model consisting of thermal plasma emission
combined with a very low photoelectric absorption provided a poor fit
($\chi^2=27.84$ for 15 degrees of freedom) and can be statistically
rejected (i.e., the probability that the best-fit model matches the
data, ${\cal Q}=0.023$, is lower than 0.05).

When we added a second plasma component to the model of the flux
during the low-level period, where both components were affected by the same
photoelectric absorption, we obtained a statistically acceptable
fit ($\chi^2=15.86$ for 13 degrees of freedom, corresponding to ${\cal
Q} = 0.26$)\footnote{The best-fit model parameters are :
a very low photoelectric absorption of $6.0\times10^{20}$~cm$^{-2}$
($1.7\times10^{20}$--$21\times10^{20}$~cm$^{-2}$ at the 90\%
confidence level); a relatively cool plasma with a temperature of
3.8~MK (2.6--6.2~MK) and an emission measure of $2.8 \times
10^{51}$~cm$^{-3}$( $1.2 \times 10^{51}$ -- $15.6 \times
10^{51}$~cm$^{-3}$); and a very hot plasma with a temperature of 794~MK
($\ge175$~MK) and an emission measure of $8.1 \times
10^{51}$~cm$^{-3}$( $5.4 \times 10^{51}$ -- $10.1\times
10^{51}$~cm$^{-3}$).}.
However, the best-fit value for the hot plasma temperature corresponds
to the highest temperature available in {\tt vapec} and the resulting lower
limit of 175~MK is too high for a PMS star \cite[e.g.,][]{preibisch05b}. 
Moreover, the residuals of the fit show a dip in the 1.2--2.0~keV
energy range and an excess in the 2.0--5.0~keV energy range,
suggesting that the modeling could be improved.
We conclude that the spectral shoulder above $\sim$1.5~keV prevents us
from obtaining any satisfactory model fit to the full spectrum during
the low-level period if the model includes a single value for the
absorbing column density for both plasma components.

Hence, to simultaneously fit both the soft and hard spectral
components, we introduced a second absorber (hydrogen column densities,
$N_\mathrm{H,s}$ and $N_\mathrm{H,h}$, for the soft and hard spectral
components, respectively). Procedurally, we first fitted the spectrum at
energies below 1.5~keV, thereby obtaining best-fit values for
$N_\mathrm{H,s}$, the plasma temperature ($T_\mathrm{s}$), and the
emission measure ($EM_\mathrm{s}$) for the soft spectral
component. Fixing these best-fit model parameters, we then fitted the whole
spectrum, and obtained best-fit values for model parameters
($N_\mathrm{H,h}$, $T_\mathrm{h}$, $EM_\mathrm{h}$) for the hard
spectral component. Using these six best-fit parameters as initial
conditions, we then fit the whole spectrum, allowing all the parameters
vary. With these initial conditions, the model parameters varied only
slightly and provided to a good fit (first line of
Table~\ref{table:fit_epic_wabs_vapec_angr_xest_parameters}).
Using the F-test, we verified that adding this second absorber
increases the goodness-of-fit at the 5\% significance level.
The soft spectral component is modeled with
lightly absorbed ($N_\mathrm{H,s}\simeq3.6\times10^{20}$~cm$^{-2}$)
emission from a 4.7~MK plasma. The hard spectral component is modeled
with emission from a 53~MK plasma affected by much larger absorption
($N_\mathrm{H,s}\simeq2.7\times10^{22}$~cm$^{-2}$). The X-ray
luminosities of the soft and the hard
components, after correcting for absorption, are $\sim4.3\times10^{28}$
and $\sim29.2\times10^{28}$~erg\,s$^{-1}$, respectively. Therefore, the
intrinsic X-ray emission of \ex~is dominated by emission from the
hotter plasma.

The bottom panel of Fig.~\ref{fig:spectra} shows the spectrum during the flaring
period, from 31.2 to 33.0~h of the whole exposure of
$\sim$6.5~ks (Fig.~\ref{fig:lc_epic}). The shape of the spectrum
during the flare incorporates both soft and hard spectral components, with
some similarity but also significant differences in overall shape from
the spectrum obtained during the low-level period. The greatest
difference between the flare and low-level spectra is the greater
prominence of the hard component during the flare.
As for the low-level emission, a model of the flare period spectrum in
which both temperature components are subject to the same absorbing column
fails. Such a model again cannot fit the data in the 1.2--2.0 and
2.0--5.0~keV energy ranges, and results in an unrealistically high value for the
temperature of the hot component. Therefore, we used a model for which
both plasma components were affected by different photoelectric
absorption when fitting the flare spectrum; however, the statistics of
the soft spectral component during the flare are very poor, due to the short flare
duration.  The weak signal from the soft component during the flare
precludes us from constraining $N_\mathrm{H,s}$ from the flare
spectrum. We therefore assumed that the absorption of the soft
spectral component did not change during the flare and we fixed the
value of $N_\mathrm{H,s}$ to that obtained from modeling the
low-level spectrum.

Our model fit to the overall flare spectrum inferred for the soft
spectral component a plasma temperature and emission measure similar
to that of the low-level period (second line of
Table~\ref{table:fit_epic_wabs_vapec_angr_xest_parameters}). For the
hard spectral component, we found a higher value for the absorption and
the plasma temperature of the hard spectral component during the flare
($5.2\times10^{22}$~cm$^{-2}$ versus $2.7\times10^{22}$~cm$^{-2}$, and
61~MK versus 53~MK, respectively).  Given the range of possible values
in our confidence interval, however, we cannot conclude with 
confidence that the absorption and the plasma temperature had increased.
We can conclude, though, that the emission measure and the intrinsic
X-ray luminosity of the hard spectral component increased by a factor
of $\sim$five during the flare.

After the source flaring period, the source faintness and the brevity
of the exposure with low-background level available after 33~h (i.e.,
$\sim$2.5~ks, see Fig.~\ref{fig:lc_flaring_background}) did not allow
us to obtain a useful spectrum.

\subsection{UV variability}
\label{UV_variability}

\begin{figure}[!ht]
\centering
\includegraphics[width=\columnwidth]{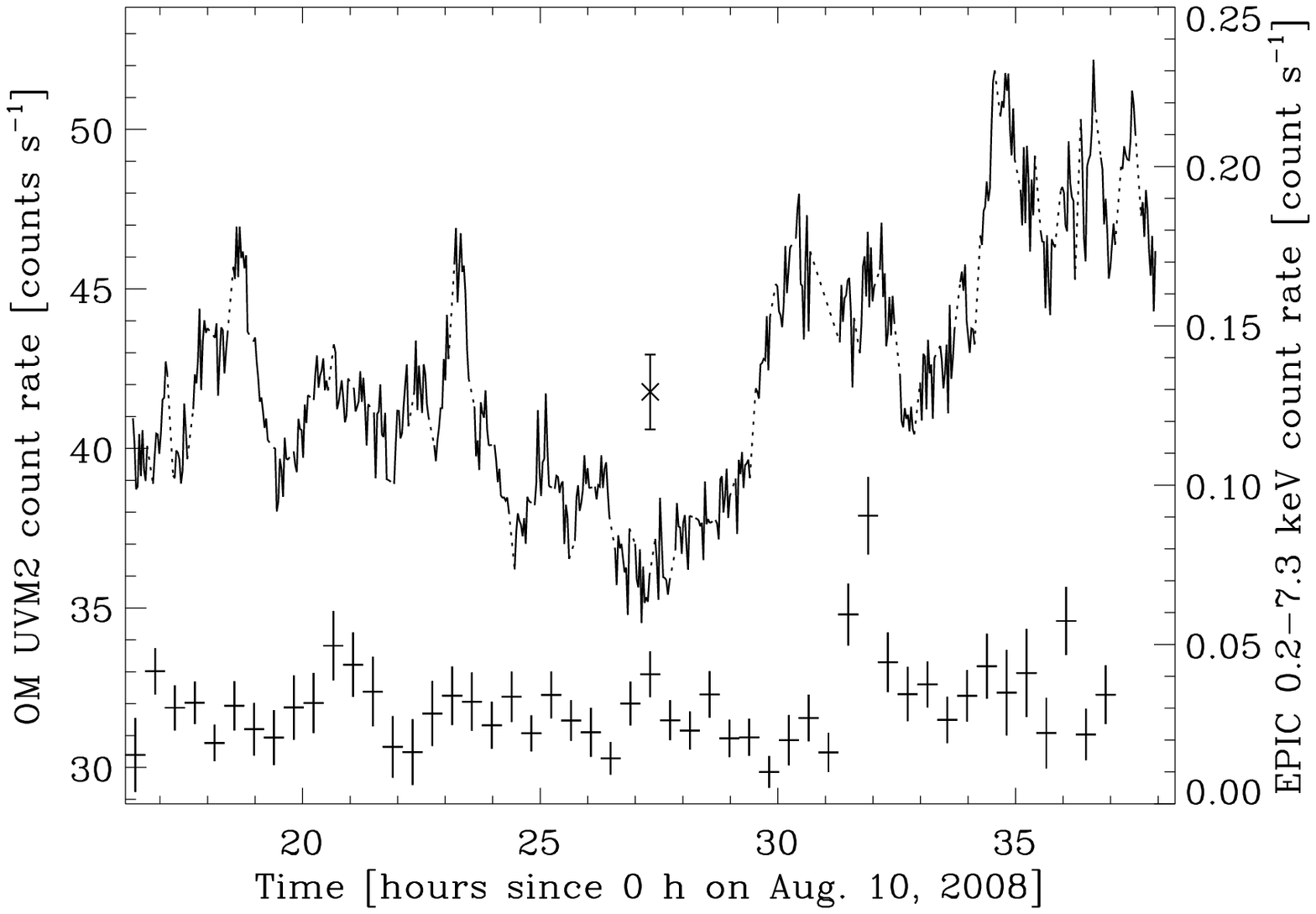}
\caption{Simultaneous UV and X-ray light curves of \ex. The upper and lower
curves are the OM light curve (left y-axis) and the EPIC light curve
(right y-axis), respectively. The UV light curve obtained with the OM
fast mode was rebinned to bin size of 120~s. In the figure center, the
cross with error bar indicates the median value and median error of the
UV photometry. To guide the eye, the dotted lines link consecutive OM
exposures. The bin size of the X-ray light curve is 500~s.
}
\label{fig:lc_om_epic}
\medskip
\includegraphics[width=\columnwidth]{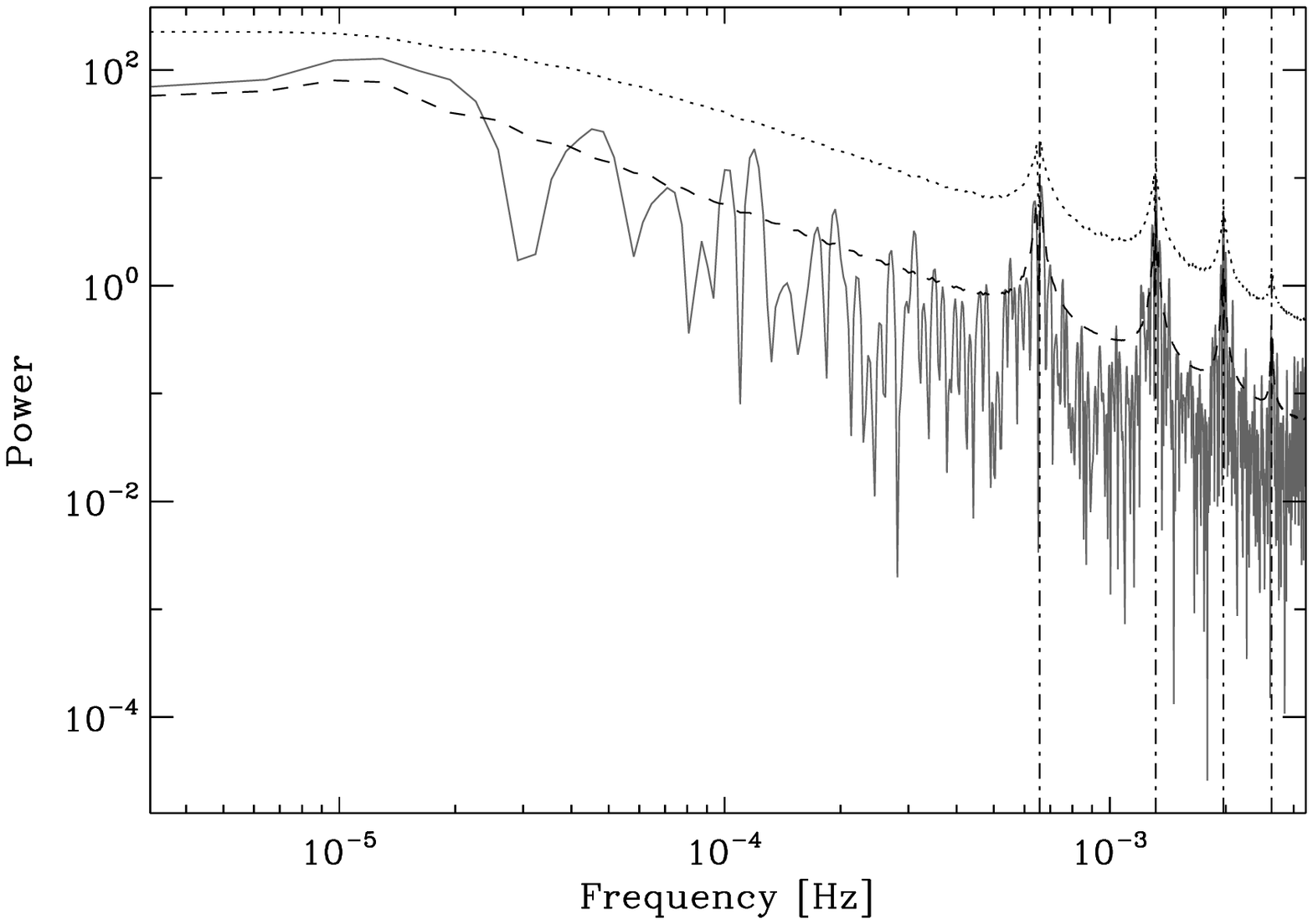}
\caption{Lomb normalized periodogram (LNP) of the UV light curve. The grey
line is the LNP of the OM light curve. The dashed and dotted
lines are the average value and the $3\sigma$ upper threshold of the LNP of red noise with power-law index of 1.39. The vertical dashed-dotted
lines indicate the frequency associated with our observing
set-up and harmonic values ($n$=2, 3, 4).}
\label{fig:psd}
\end{figure}

The UV event lists produced by the OM fast mode were rebinned to a bin
size of 120~s to increase the signal-to-noise ratio. We obtained 10 UV
measurements per OM fast-mode exposure, thus, 500 UV measurements for
the whole observation. Figure~\ref{fig:lc_om_epic} shows this UV light
curve and, for comparison, the X-ray light curve.
Significant UV variability was detected, with several events
lasting from shorter than one hour to as long as about four hours.
During these one-to-four hour events, the UV flux increased by ten to
twenty percent from the initial count rate before returning to it.
The baseline level of the UV light curve was not constant. In the first
half of the observation, it followed a downward trend from about
40~counts\,s$^{-1}$ to a minimum of about 36~counts\,s$^{-1}$, and,
then, an upward trend to a maximum of about 45~counts\,s$^{-1}$.

None of the multiple UV peaks correlated simultaneously with any of
the possible changes in the X-ray count rate. The rise of the X-ray
flare occurred about 2~h after the start of a large UV flare.
However, we cannot conclude, definitively, that absolutely no
connection exists between the UV and X-ray flare events.

To search for any periodicity in this UV light
curve, we computed the Lomb normalized periodogram
\citep[LNP;][]{lomb76,scargle82,press89}, which is well suited to unevenly
sampled time series. The LNP revealed that the power of the UV signal
versus frequency has a power-law shape (i.e., red noise), in addition
to several high frequency peaks (Fig.~\ref{fig:psd}). A least squares
log-log fitting of the LNP
versus frequency inferred a negative slope,
$a=-1.26\pm0.05$. However, our observing procedure may flatten the LNP at
high frequencies \citep[e.g.,][]{uttley02},
hence the value of the LNP slope is not simply the negative
of the power-law index ($\alpha$) of the true power spectral density (PSD),
$PSD(f)\propto f^{-\alpha}$, where $f$ is the frequency (see our
simulations below). Nevertheless, this best-fit value of $a$ indicates that
$\alpha$ is most likely to be between 1 and 2, which are values of $\alpha$
associated with flicker noise 
and random-walk noise \citep{press78}, respectively. In contrast to
the case of Gaussian
white noise ($\alpha=0$), where the significance level of any peak in
the LNP can be evaluated analytically \citep{scargle82}, we needed here
to perform detailed simulations to assess the level of significance of
the high frequency peaks.

We assumed first that the true PSD of the
intrinsic UV time series is that of red noise that we sampled evenly at the
Nyquist frequency of our observation ($f_\mathrm{\,Nyquist}=2/120$~Hz)
from 60~s to 10 times the duration of our observation, and we then used
the algorithm of \cite{timmer95} to produce the corresponding light curve. 
We divided this light
curve into 10 shorter light curves matching the duration of our
observation. Each light curve was then binned in the same way as our UV
light curve, and scaled to match the UV mean count rate and standard
deviation. The corresponding LNPs were then computed, leading to a
sample of 10 LNP. We repeated this procedure 10\,000 times to obtain a large
(100\,000) sample of LNPs. 
We found that red noise with $\alpha=1.39\pm0.06$
reproduces the slope of the LNP of the observed UV light curve.
The observed high frequency peaks are below the
upper threshold of the 99.7\% confidence level (equivalent to $3\sigma$ in
the Gaussian case) of the random fluctuation of the LNP, and are, hence,
not significant. Indeed, our simulated light curves display similar
high frequency peaks that we identify as aliasing produced by our
observing windows. The lower frequency ($f \approx
6.6\times10^{-4}$~Hz~$\approx 1/1522$~Hz)
matches the frequency associated with the sum of fast mode exposure
(1200~s) and exposure gap (321.6~s); the other frequencies are harmonics
($n \times f$, with $n$=2, 3, 4).

We conclude that the UV variability can be
modeled with red noise ($\alpha=1.39\pm0.06$) that arises from a
true variability of the underlying physical process. We detected no
significant, periodic signal in the UV light curve.

\section{Discussion}
\label{discussion}

\subsection{Two-absorber X-ray sources}

The X-ray spectrum of \ex~exhibits two plasma components, one that is cooler
with low absorption and one that is hotter with high absorption. Similar
soft and hard spectral components with two absorbers have previously
been observed in a few low-mass stars and protostars with accretion disks 
and jets that have been detected in forbidden emission lines
\citep{kastner05,guedel05,guedel07d}. 

The acronym TAX, which stands for Two-Absorber X-ray, was introduced
by \cite{guedel07d} to classify jet-driving sources that display
this peculiar X-ray spectrum. The interpretation of the X-ray
spectra of young stellar objects classified as TAX sources is that
variable emission from the hotter plasma of the stellar corona is
absorbed by circumstellar material, producing the hard spectral component. The
{\sl non-variable}, lightly absorbed soft component, meanwhile, likely arises
as a consequence of shocks located at the base of jet-launching
zones.

The prototype of the FUor class, FU~Ori, shows this
kind of unusual X-ray spectrum, but the $\sim 8$~MK temperature of
the soft component in FU~Ori may reflect a coronal magnetic
process rather than a shock origin \citep{skinner06}.

\subsection{The origin of the hard X-ray component of \ex}
\label{origin_hard}

The flaring activity of the hard component of \ex, combined with the
hotter plasma temperature of this component, points clearly to
magnetic processes and probably coronal activity as the source of
these X-rays.
For comparison, the fractional X-ray activity\footnote{Following
\cite{kenyon95}, we compute the stellar
luminosity of \ex, $L_{\rm bol}$, from its magnitude in the $J$-band
-- specifically 2MASS (low-level) measurement of 9.73\,mag, transformed
to the homogenized system of \cite{bessell88} using the color
transformations of \cite{carpenter01} -- using $V-J$ color and bolometric
correction appropriate for a M0 spectral type star \citep[see Table~A5 of
][]{kenyon95}. We obtain $L_{\rm bol}=0.55~L_\odot$, and a
corresponding stellar radius of 1.7~$R_\odot$ for an effective
temperature of 3850~K \citep{kenyon95}.} of \ex, during the
low-level period, $\log(L_\mathrm{X}/L_{\rm
bol})=-3.8$, is similar to the median value observed
for (accreting) T~Tauri stars of the Orion nebula cluster \citep[i.e.,
$-3.74$;][]{preibisch05b}.
Moreover, the flare amplitude is well within the range of values observed 
in this sample for T~Tauri stars having similar value of low-level X-ray
luminosity \citep[see
Fig.~3b of][]{stassun07}.

The large photoelectric absorption of the hotter plasma component of \ex~is
equivalent to an optical extinction of $\sim15$~mag
\citep{vuong03}. However, the value of the optical extinction
of \ex, although not well known, is very likely much lower, and generally
{\sl assumed} to be 0 or 1~mag in the literature
\citep[e.g.,][]{herbig01,gras-velazquez05,sipos09}. \cite{herbig01}
noticed that the near-IR colors of \ex~obtained in March 1992
\citep{hughes94} are redder than those of normal M0 dwarfs. 
We found here that the reddened
color of \ex~in the near-IR is consistent with the intrinsic
near-IR excesses of classical T~Tauri stars (CTTSs), which are caused
by direct emission of near-IR flux from disks and not by the
reddening of photospheric emission by dust. In particular,
2MASS near-IR colors of \ex~($J-H=0.770\pm0.032$~mag,
$H-K=0.462\pm0.030$~mag) -- obtained on May 16, 1999, 
when \ex~was at its minimum level \citep[a visual
magnitude of 13.5;][]{herbig07} -- are well located on the loci of
CTTSs \citep{meyer97} tranformed to the 2MASS color system
\cite[using the color transformations of][]{carpenter01}. Therefore,
we conclude that the optical extinction of \ex~is fully consistent
with the argument that \ex~experiences no optical extinction ($A_{\mathrm
V}=0$~mag) by dusty material.
Moreover, \cite{sipos09} argued, from modeling of the spectral energy
distribution of \ex~during quiescence, that the accretion disk is
viewed more nearly face-on than edge-on, with a best-fit model
inclination of $20\degr$. This low inclination prevents any
absorption of the stellar coronal emission by circumstellar
material (gas+dust), which is consistent with the null optical
extinction. 

The large column density is hence necessarily produced by
a dust-free gas with a low ionization. The accreted material close to a few
stellar radii of the stellar photosphere is naturally dust-free after the
sublimation of the dust grains. Moreover, the accretion funnels,
loaded with high density gas ($\sim10^{12}$~cm$^{-3}$), can easily
produce this large column density on a length scale of only 0.3 stellar
radii. The more or less pole-on view may also favor the absorption
by these accretion funnels of X-ray emission from stellar magnetic
loops at lower latitudes \citep{gregory07}.

On the surface, the sustained presence of a luminous, hard (coronal)
X-ray component toward the end of the outburst of \ex~would seem to be
at odds with models in which energetic ions emitted during coronal flares
transform disk dust grains from silicate to amorphous form \citep{glauser09},
given that the reverse was observed during the \ex~outburst
\citep{abraham09}. We speculate that the large column density observed
in front of a coronal active region during the 2008 extreme outburst of
\ex~protected the fresh crystalline silicate apparently produced, against
amorphization, hence possibly encouraging the production of cometary material
during accretion outburst.  Further modeling of the potential
effects of PMS outbursts and flares on disk dust-grain structure is
clearly warranted.

\subsection{The origin of the soft X-ray component of \ex}

The low plasma temperature of the soft component is consistent with
X-ray emission from a shock, located either inside a jet, as
proposed for TAX sources \citep{kastner05,guedel05,guedel07d},
or at the base of an accretion shock, as suggested by analyses of
high-resolution X-ray spectra of CTTSs
\citep{kastner02,stelzer04c,schmitt05,guenther06,argiroffi07,robrade07b,brickhouse10}.

The former is unlikely as there is no indication of a jet in \ex.
Forbidden emission lines of [\ion{O}{i}], [\ion{N}{ii}],
and [\ion{S}{ii}], which are the jet signatures visible in optical spectrum of CTTSs
\citep[e.g.,][]{hirth97}, are not detected in the optical spectra of
\ex~(e.g., \citealt{kospal08}, and references therein; see
in particular \citealt{herbig07} for Keck/HIRES optical spectrum
showing a lack of [\ion{S}{ii}] emission lines).

Furthermore, in contrast to TAX sources
(e.g., Fig.~2 of \citealt{guedel07d}), variability is evident in the
soft component of the \ex~X-ray spectrum. Hence, we favor the latter
interpretation of the soft X-ray component, i.e., that this component has
its origin in accretion shocks. As in the case of TW~Hya
\citep{kastner02,brickhouse10}, our pole-on view of the \ex~star-disk
system indeed probably provides a direct view of the relatively cool plasma
associated with these shocks.

\subsection{The origin of the UV emission of \ex}

The UV light curve of \ex~is reminiscent of the UV activity already
observed with the OM from other accreting pre-main sequence stars
\citep[e.g., BP~Tau;][]{schmitt05}. 
In these accreting systems, the UV emission is
directly related to a hot spot on the stellar surface
\citep{calvet98,gullbring98}. In this framework of magnetospheric
accretion \citep[see review of ][]{bouvier07b}, hot spots are
located at the base of magnetic loops, which connect the photosphere and
the accretion disk and are loaded with free-falling material
that produces accretion shocks at the stellar surface. 

Following \cite{schmitt05}, we estimated the surface filling factor of the
hot spots, i.e., the ratio of the hot spot area to the surface of
the \ex~photosphere. Using the UVM2 filter  count rate to flux
conversion factor\footnote{See
\href{http://xmm2.esac.esa.int/docs/documents/CAL-TN-0019.ps.gz}{the
OM calibration status (issue 5.0)}.} of 
$2.19 \times 10^{-15}$~erg\,cm$^{-2}$\,s$^{-1}$\,$\AA^{-1}$, we
converted the minimum count rate of $\sim$36~counts\,s$^{-1}$ to
$1.8\times10^{-10}$~erg\,cm$^2$\,s$^{-1}$. As we discussed
in Sect.~\ref{origin_hard}, the optical extinction is negligible in
\ex,  therefore, we could consider this value to be the intrinsic UV
flux. We assumed for the hot spot a black-body spectrum of temperature
13\,000~K, as measured
from an {\sl IUE} spectrum obtained during the 1994
outburst, when \ex~was at $V=12.1$~mag \citep{herbig01}, i.e., 3.3
times fainter than during our observation. We found a
surface filling factor of 0.9\% for the hot spot. This
small surface-filling factor is typical of values found in CTTSs 
\citep{calvet98}.

The increase by about 45\% in the count rate from the minimum to the
maximum observed value over a period of about 10~h
can be explained either by an increase in the surface filling
factor of the hot spot in the same proportion, or by an increase of
1000~K in the hot spot temperature. The latter can be produced by
an increase in the mass accretion rate of a factor of $\sim$1.2
\citep[an estimate we obtained by interpolating the grid of models
of ][]{calvet98}.

\section{Conclusions}
\label{conclusions}

We observed \ex~with \xmm~a few days before the end of its 2008 extreme
optical outburst, but when \ex~was still in an elevated state. This
observation is the longest continuous exposure that was obtained in
X-rays and UV during this optical outburst. 

We have found that the X-ray light curve of
\ex~contains mainly a low-level period and a two-hour flare event,
during which the observed flux increased by a
factor of four and noticeably hardened. 
The variability observed in the UV on timescales ranging from less than
one hour up to about four hours can be modeled with a red-noise
spectrum. None of the UV events observed on August 10--11, 2008
correlate unambiguously with simultaneous X-ray peaks. This UV activity is
typical of accretion events and is dominated by emission from
accretion hot spots covering only about one percent of the stellar surface.

The X-ray spectrum of \ex~reveals two plasma components, one that is cooler
with low absorption and one that is hotter with high absorption. The
intrinsic X-ray luminosity of \ex~is dominated by the emission from
the hot plasma component. During the X-ray flare, the emission measure
and the intrinsic X-ray luminosity of this absorbed plasma component
is five times greater.

The soft plasma component is most likely associated with X-ray
emission from accretion shocks -- an inference based mainly on 
the absence of a jet in \ex. 
The nearly
pole-on viewing geometry likely explains the
high absorption of X-ray emission from low-latitude active regions by the
high-density accretion funnel flows. Hence, the same dust-free
accreting material that generates the soft component (via shocks) also
largely smothers the corona of \ex. This model predicts that
large values of column density, such as we measured toward the hard spectral
component of \ex~near the end of its 2008 outburst, should be present
only during these accretion-driven optical outbursts.

\begin{acknowledgements}
We thanks the \xmm~Science Operations Centre for the prompt schedule
of this observation. This research is based on observations obtained
with \xmm, an ESA science mission with instruments and contributions
directly funded by ESA Member States and NASA. We acknowledge with
thanks the variable star observations from the AAVSO International
Database contributed by observers worldwide and used in this research.
\end{acknowledgements}

\phantomsection
\addcontentsline{toc}{chapter}{References}
\bibliographystyle{aa}
\bibliography{/home/grosso/references/references}

\begin{thebibliography}{74}
\expandafter\ifx\csname natexlab\endcsname\relax\def\natexlab#1{#1}\fi

\bibitem[{{{\'A}brah{\'a}m} {et~al.}(2009){{\'A}brah{\'a}m}, {Juh{\'a}sz},
  {Dullemond}, {K{\'o}sp{\'a}l}, {van Boekel}, {Bouwman}, {Henning},
  {Mo{\'o}r}, {Mosoni}, {Sicilia-Aguilar}, \& {Sipos}}]{abraham09}
{{\'A}brah{\'a}m}, P., {Juh{\'a}sz}, A., {Dullemond}, C.~P., {et~al.} 2009,
  \nat, 459, 224

\bibitem[{{Anders} \& {Ebihara}(1982)}]{anders82}
{Anders}, E. \& {Ebihara}, M. 1982, \gca, 46, 2363

\bibitem[{{Anders} \& {Grevesse}(1989)}]{anders89}
{Anders}, E. \& {Grevesse}, N. 1989, \gca, 53, 197

\bibitem[{{Argiroffi} {et~al.}(2007){Argiroffi}, {Maggio}, \&
  {Peres}}]{argiroffi07}
{Argiroffi}, C., {Maggio}, A., \& {Peres}, G. 2007, \aap, 465, L5

\bibitem[{{Audard} {et~al.}(2005){Audard}, {G{\"u}del}, {Skinner}, {Briggs},
  {Walter}, {Stringfellow}, {Hamilton}, \& {Guinan}}]{audard05b}
{Audard}, M., {G{\"u}del}, M., {Skinner}, S.~L., {et~al.} 2005, \apjl, 635, L81

\bibitem[{{Audard} {et~al.}(2010){Audard}, {Stringfellow}, {G{\"u}del},
  {Skinner}, {Walter}, {Guinan}, {Hamilton}, {Briggs}, \&
  {Baldovin-Saavedra}}]{audard10}
{Audard}, M., {Stringfellow}, G.~S., {G{\"u}del}, M., {et~al.} 2010, \aap, 511,
  A63

\bibitem[{{Bessell} \& {Brett}(1988)}]{bessell88}
{Bessell}, M.~S. \& {Brett}, J.~M. 1988, \pasp, 100, 1134

\bibitem[{{Bouvier} {et~al.}(2007){Bouvier}, {Alencar}, {Harries},
  {Johns-Krull}, \& {Romanova}}]{bouvier07b}
{Bouvier}, J., {Alencar}, S.~H.~P., {Harries}, T.~J., {Johns-Krull}, C.~M., \&
  {Romanova}, M.~M. 2007, in Protostars and Planets V, ed. B.~{Reipurth},
  D.~{Jewitt}, \& K.~{Keil}, 479--494

\bibitem[{{Brickhouse} {et~al.}(2010){Brickhouse}, {Cranmer}, {Dupree}, {Luna},
  \& {Wolk}}]{brickhouse10}
{Brickhouse}, N.~S., {Cranmer}, S.~R., {Dupree}, A.~K., {Luna}, G.~J.~M., \&
  {Wolk}, S. 2010, \apj, 710, 1835

\bibitem[{{Calvet} \& {Gullbring}(1998)}]{calvet98}
{Calvet}, N. \& {Gullbring}, E. 1998, \apj, 509, 802

\bibitem[{{Carpenter}(2001)}]{carpenter01}
{Carpenter}, J.~M. 2001, \aj, 121, 2851

\bibitem[{{Comer{\'o}n}(2008)}]{comeron08}
{Comer{\'o}n}, F. 2008, The Lupus Clouds, in Handbook of Star Forming Regions,
  Volume II: The Southern Sky, ASP Monograph Publications, Vol.~5, ed.
  B.~Reipurth, p.~295--350

\bibitem[{{Dorman} \& {Arnaud}(2001)}]{dorman01}
{Dorman}, B. \& {Arnaud}, K.~A. 2001, in Astronomical Society of the Pacific
  Conference Series, Vol. 238, Astronomical Data Analysis Software and Systems
  X, ed. .~H.~E.~P. F.~R.~Harnden~Jr., F.~A.~Primini, 415

\bibitem[{{Feigelson} \& {Montmerle}(1999)}]{feigelson99}
{Feigelson}, E.~D. \& {Montmerle}, T. 1999, \araa, 37, 363

\bibitem[{{G{\" u}del} {et~al.}(2005){G{\" u}del}, {Skinner}, {Briggs},
  {Audard}, {Arzner}, \& {Telleschi}}]{guedel05}
{G{\" u}del}, M., {Skinner}, S.~L., {Briggs}, K.~R., {et~al.} 2005, \apjl, 626,
  L53

\bibitem[{{Getman} {et~al.}(2005){Getman}, {Flaccomio}, {Broos}, {Grosso},
  {Tsujimoto}, {Townsley}, {Garmire}, {Kastner}, {Li}, {Harnden}, {Wolk},
  {Murray}, {Lada}, {Muench}, {McCaughrean}, {Meeus}, {Damiani}, {Micela},
  {Sciortino}, {Bally}, {Hillenbrand}, {Herbst}, {Preibisch}, \&
  {Feigelson}}]{getman05}
{Getman}, K.~V., {Flaccomio}, E., {Broos}, P.~S., {et~al.} 2005, \apjs, 160,
  319

\bibitem[{{Glauser} {et~al.}(2009){Glauser}, {G{\"u}del}, {Watson}, {Henning},
  {Schegerer}, {Wolf}, {Audard}, \& {Baldovin-Saavedra}}]{glauser09}
{Glauser}, A.~M., {G{\"u}del}, M., {Watson}, D.~M., {et~al.} 2009, \aap, 508,
  247

\bibitem[{{Gras-Vel{\'a}zquez} \& {Ray}(2005)}]{gras-velazquez05}
{Gras-Vel{\'a}zquez}, {\`A}. \& {Ray}, T.~P. 2005, \aap, 443, 541

\bibitem[{{Gregory} {et~al.}(2007){Gregory}, {Wood}, \& {Jardine}}]{gregory07}
{Gregory}, S.~G., {Wood}, K., \& {Jardine}, M. 2007, \mnras, 379, L35

\bibitem[{{Grosso}(2006)}]{grosso06c}
{Grosso}, N. 2006, in ESA Special Publication, Vol. 604, The X-ray Universe
  2005, ed. A.~{Wilson}, 51--56

\bibitem[{{Grosso} {et~al.}(2005){Grosso}, {Kastner}, {Ozawa}, {Richmond},
  {Simon}, {Weintraub}, {Hamaguchi}, \& {Frank}}]{grosso05}
{Grosso}, N., {Kastner}, J.~H., {Ozawa}, H., {et~al.} 2005, \aap, 438, 159

\bibitem[{{G{\"u}del} {et~al.}(2007{\natexlab{a}}){G{\"u}del}, {Briggs},
  {Arzner}, {Audard}, {Bouvier}, {Feigelson}, {Franciosini}, {Glauser},
  {Grosso}, {Micela}, {Monin}, {Montmerle}, {Padgett}, {Palla}, {Pillitteri},
  {Rebull}, {Scelsi}, {Silva}, {Skinner}, {Stelzer}, \& {Telleschi}}]{guedel07}
{G{\"u}del}, M., {Briggs}, K.~R., {Arzner}, K., {et~al.} 2007{\natexlab{a}},
  \aap, 468, 353

\bibitem[{{G{\"u}del} \& {Naz{\'e}}(2009)}]{guedel09}
{G{\"u}del}, M. \& {Naz{\'e}}, Y. 2009, \aapr, 7

\bibitem[{{G{\"u}del} {et~al.}(2007{\natexlab{b}}){G{\"u}del}, {Telleschi},
  {Audard}, {L.~Skinner}, {Briggs}, {Palla}, \& {Dougados}}]{guedel07d}
{G{\"u}del}, M., {Telleschi}, A., {Audard}, M., {et~al.} 2007{\natexlab{b}},
  \aap, 468, 515

\bibitem[{{Gullbring} {et~al.}(1998){Gullbring}, {Hartmann}, {Brice\~no}, \&
  {Calvet}}]{gullbring98}
{Gullbring}, E., {Hartmann}, L., {Brice\~no}, C., \& {Calvet}, N. 1998, \apj,
  492, 323

\bibitem[{{G{\"u}nther} {et~al.}(2006){G{\"u}nther}, {Liefke}, {Schmitt},
  {Robrade}, \& {Ness}}]{guenther06}
{G{\"u}nther}, H.~M., {Liefke}, C., {Schmitt}, J.~H.~M.~M., {Robrade}, J., \&
  {Ness}, J.-U. 2006, \aap, 459, L29

\bibitem[{{Hamaguchi} {et~al.}(2010){Hamaguchi}, {Grosso}, {Kastner},
  {Weintraub}, \& {Richmond}}]{hamaguchi10}
{Hamaguchi}, K., {Grosso}, N., {Kastner}, J.~H., {Weintraub}, D.~A., \&
  {Richmond}, M. 2010, \apjl, 714, L16

\bibitem[{{Hartmann} \& {Kenyon}(1996)}]{hartmann96}
{Hartmann}, L. \& {Kenyon}, S.~J. 1996, \araa, 34, 207

\bibitem[{{Herbig}(1966)}]{herbig66}
{Herbig}, G.~H. 1966, Vistas in Astronomy, 8, 109

\bibitem[{{Herbig}(1977{\natexlab{a}})}]{herbig77}
{Herbig}, G.~H. 1977{\natexlab{a}}, \apj, 217, 693

\bibitem[{{Herbig}(1977{\natexlab{b}})}]{herbig77b}
{Herbig}, G.~H. 1977{\natexlab{b}}, \apj, 214, 747

\bibitem[{{Herbig}(1989)}]{herbig89}
{Herbig}, G.~H. 1989, in European Southern Observatory Astrophysics Symposia,
  Vol.~33, ESO Workshop on Low Mass Star Formation and Pre-Main Sequence
  Objects, ed. B.~Reipurth, p.~233--246

\bibitem[{{Herbig}(2007)}]{herbig07}
{Herbig}, G.~H. 2007, \aj, 133, 2679

\bibitem[{{Herbig}(2008)}]{herbig08}
{Herbig}, G.~H. 2008, \aj, 135, 637

\bibitem[{{Herbig} {et~al.}(2001){Herbig}, {Aspin}, {Gilmore}, {Imhoff}, \&
  {Jones}}]{herbig01}
{Herbig}, G.~H., {Aspin}, C., {Gilmore}, A.~C., {Imhoff}, C.~L., \& {Jones},
  A.~F. 2001, \pasp, 113, 1547

\bibitem[{{Hirth} {et~al.}(1997){Hirth}, {Mundt}, \& {Solf}}]{hirth97}
{Hirth}, G.~A., {Mundt}, R., \& {Solf}, J. 1997, \aaps, 126, 437

\bibitem[{{Hughes} {et~al.}(1994){Hughes}, {Hartigan}, {Krautter}, \&
  {Kelemen}}]{hughes94}
{Hughes}, J., {Hartigan}, P., {Krautter}, J., \& {Kelemen}, J. 1994, \aj, 108,
  1071

\bibitem[{{Itagaki} {et~al.}(2008){Itagaki}, {Nakano}, \&
  {Yamaoka}}]{itagaki08}
{Itagaki}, K., {Nakano}, S., \& {Yamaoka}, H. 2008, \iaucirc, 8968, 2

\bibitem[{{Jones}(2008{\natexlab{a}})}]{jones08}
{Jones}, A.~F. 2008{\natexlab{a}}, Central Bureau Electronic Telegrams, 1217, 1

\bibitem[{{Jones}(2008{\natexlab{b}})}]{jones08b}
{Jones}, A.~F. 2008{\natexlab{b}}, Central Bureau Electronic Telegrams, 1231, 1

\bibitem[{{Kastner} {et~al.}(2006{\natexlab{a}}){Kastner}, {Richmond}, {Simon},
  {Grosso}, {Weintraub}, \& {Hamaguchi}}]{kastner06b}
{Kastner}, J., {Richmond}, M., {Simon}, T., {et~al.} 2006{\natexlab{a}},
  Central Bureau Electronic Telegrams, 760, 1

\bibitem[{{Kastner} {et~al.}(2005){Kastner}, {Franz}, {Grosso}, {Bally},
  {McCaughrean}, {Getman}, {Feigelson}, \& {Schulz}}]{kastner05}
{Kastner}, J.~H., {Franz}, G., {Grosso}, N., {et~al.} 2005, \apjs, 160, 511

\bibitem[{{Kastner} {et~al.}(2002){Kastner}, {Huenemoerder}, {Schulz},
  {Canizares}, \& {Weintraub}}]{kastner02}
{Kastner}, J.~H., {Huenemoerder}, D.~P., {Schulz}, N.~S., {Canizares}, C.~R.,
  \& {Weintraub}, D.~A. 2002, \apj, 567, 434

\bibitem[{{Kastner} {et~al.}(2004){Kastner}, {Richmond}, {Grosso}, {Weintraub},
  {Simon}, {Frank}, {Hamaguchi}, {Ozawa}, \& {Henden}}]{kastner04}
{Kastner}, J.~H., {Richmond}, M., {Grosso}, N., {et~al.} 2004, \nat, 430, 429

\bibitem[{{Kastner} {et~al.}(2006{\natexlab{b}}){Kastner}, {Richmond},
  {Grosso}, {Weintraub}, {Simon}, {Henden}, {Hamaguchi}, {Frank}, \&
  {Ozawa}}]{kastner06}
{Kastner}, J.~H., {Richmond}, M., {Grosso}, N., {et~al.} 2006{\natexlab{b}},
  \apjl, 648, L43

\bibitem[{{Kenyon} \& {Hartmann}(1995)}]{kenyon95}
{Kenyon}, S.~J. \& {Hartmann}, L. 1995, \apjs, 101, 117

\bibitem[{{K\'osp\'al} {et~al.}(2008){K\'osp\'al}, {Nemeth}, {Abraham}, {Kun},
  {Henden}, \& {Jones}}]{kospal08}
{K\'osp\'al}, A., {Nemeth}, P., {Abraham}, P., {et~al.} 2008, Information
  Bulletin on Variable Stars, 5819, 1

\bibitem[{{Lomb}(1976)}]{lomb76}
{Lomb}, N.~R. 1976, \apss, 39, 447

\bibitem[{{Lombardi} {et~al.}(2008){Lombardi}, {Lada}, \& {Alves}}]{lombardi08}
{Lombardi}, M., {Lada}, C.~J., \& {Alves}, J. 2008, \aap, 480, 785

\bibitem[{{Lorenzetti} {et~al.}(2006){Lorenzetti}, {Giannini}, {Calzoletti},
  {Puccetti}, {Antoniucci}, {Arkharov}, {di Paola}, {Larionov}, \&
  {Nisini}}]{lorenzetti06}
{Lorenzetti}, D., {Giannini}, T., {Calzoletti}, L., {et~al.} 2006, \aap, 453,
  579

\bibitem[{{Mason} {et~al.}(2001){Mason}, {Breeveld}, {Much}, {Carter},
  {Cordova}, {Cropper}, {Fordham}, {Huckle}, {Ho}, {Kawakami}, {Kennea},
  {Kennedy}, {Mittaz}, {Pandel}, {Priedhorsky}, {Sasseen}, {Shirey}, {Smith},
  \& {Vreux}}]{mason01}
{Mason}, K.~O., {Breeveld}, A., {Much}, R., {et~al.} 2001, \aap, 365, L36

\bibitem[{{Meyer} {et~al.}(1997){Meyer}, {Calvet}, \& {Hillenbrand}}]{meyer97}
{Meyer}, M.~R., {Calvet}, N., \& {Hillenbrand}, L.~A. 1997, \aj, 114, 288

\bibitem[{{Morrison} \& {McCammon}(1983)}]{morrison83}
{Morrison}, R. \& {McCammon}, D. 1983, \apj, 270, 119

\bibitem[{{Pojmanski}(2002)}]{pojmanski02}
{Pojmanski}, G. 2002, Acta Astronomica, 52, 397

\bibitem[{{Preibisch} {et~al.}(2005){Preibisch}, {Kim}, {Favata}, {Feigelson},
  {Flaccomio}, {Getman}, {Micela}, {Sciortino}, {Stassun}, {Stelzer}, \&
  {Zinnecker}}]{preibisch05b}
{Preibisch}, T., {Kim}, Y.-C., {Favata}, F., {et~al.} 2005, \apjs, 160, 401

\bibitem[{{Press}(1978)}]{press78}
{Press}, W.~H. 1978, Comments on Astrophysics, 7, 103

\bibitem[{{Press} \& {Rybicki}(1989)}]{press89}
{Press}, W.~H. \& {Rybicki}, G.~B. 1989, \apj, 338, 277

\bibitem[{{Robrade} \& {Schmitt}(2007)}]{robrade07b}
{Robrade}, J. \& {Schmitt}, J.~H.~M.~M. 2007, \aap, 473, 229

\bibitem[{{Scargle}(1982)}]{scargle82}
{Scargle}, J.~D. 1982, \apj, 263, 835

\bibitem[{{Schmitt} {et~al.}(2005){Schmitt}, {Robrade}, {Ness}, {Favata}, \&
  {Stelzer}}]{schmitt05}
{Schmitt}, J.~H.~M.~M., {Robrade}, J., {Ness}, J.-U., {Favata}, F., \&
  {Stelzer}, B. 2005, \aap, 432, L35

\bibitem[{{Sipos} {et~al.}(2009){Sipos}, {{\'A}brah{\'a}m}, {Acosta-Pulido},
  {Juh{\'a}sz}, {K{\'o}sp{\'a}l}, {Kun}, {Mo{\'o}r}, \& {Setiawan}}]{sipos09}
{Sipos}, N., {{\'A}brah{\'a}m}, P., {Acosta-Pulido}, J., {et~al.} 2009, \aap,
  507, 881

\bibitem[{{Skinner} {et~al.}(2006){Skinner}, {Briggs}, \&
  {G{\"u}del}}]{skinner06}
{Skinner}, S.~L., {Briggs}, K.~R., \& {G{\"u}del}, M. 2006, \apj, 643, 995

\bibitem[{{Skinner} {et~al.}(2009){Skinner}, {Sokal}, {G{\"u}del}, \&
  {Briggs}}]{skinner09}
{Skinner}, S.~L., {Sokal}, K.~R., {G{\"u}del}, M., \& {Briggs}, K.~R. 2009,
  \apj, 696, 766

\bibitem[{{Stassun} {et~al.}(2007){Stassun}, {van den Berg}, \&
  {Feigelson}}]{stassun07}
{Stassun}, K.~G., {van den Berg}, M., \& {Feigelson}, E. 2007, \apj, 660, 704

\bibitem[{{Stecklum}(2006)}]{stecklum06}
{Stecklum}, B. 2006, Central Bureau Electronic Telegrams, 690, 1

\bibitem[{{Stecklum} {et~al.}(2007){Stecklum}, {Melnikov}, \&
  {Meusinger}}]{stecklum07}
{Stecklum}, B., {Melnikov}, S.~Y., \& {Meusinger}, H. 2007, \aap, 463, 621

\bibitem[{{Stelzer} {et~al.}(2009){Stelzer}, {Hubrig}, {Orlando}, {Micela},
  {Mikul{\'a}{\v s}ek}, \& {Sch{\"o}ller}}]{stelzer09}
{Stelzer}, B., {Hubrig}, S., {Orlando}, S., {et~al.} 2009, \aap, 499, 529

\bibitem[{{Stelzer} \& {Schmitt}(2004)}]{stelzer04c}
{Stelzer}, B. \& {Schmitt}, J.~H.~M.~M. 2004, \aap, 418, 687

\bibitem[{{Str{\"u}der} {et~al.}(2001){Str{\"u}der}, {Briel}, {Dennerl},
  {Hartmann}, {Kendziorra}, {Meidinger}, {Pfeffermann}, {Reppin}, {Aschenbach},
  {Bornemann}, {Br{\"a}uninger}, {Burkert}, {Elender}, {Freyberg}, {Haberl},
  {Hartner}, {Heuschmann}, {Hippmann}, {Kastelic}, {Kemmer}, {Kettenring},
  {Kink}, {Krause}, {M{\"u}ller}, {Oppitz}, {Pietsch}, {Popp}, {Predehl},
  {Read}, {Stephan}, {St{\"o}tter}, {Tr{\"u}mper}, {Holl}, {Kemmer}, {Soltau},
  {St{\"o}tter}, {Weber}, {Weichert}, {von Zanthier}, {Carathanassis}, {Lutz},
  {Richter}, {Solc}, {B{\"o}ttcher}, {Kuster}, {Staubert}, {Abbey}, {Holland},
  {Turner}, {Balasini}, {Bignami}, {La Palombara}, {Villa}, {Buttler},
  {Gianini}, {Lain{\'e}}, {Lumb}, \& {Dhez}}]{strueder01}
{Str{\"u}der}, L., {Briel}, U., {Dennerl}, K., {et~al.} 2001, \aap, 365, L18

\bibitem[{{Szeifert} {et~al.}(2010){Szeifert}, {Hubrig}, {Sch{\"o}ller},
  {Sch{\"u}tz}, {Stelzer}, \& {Mikul{\'a}{\v s}ek}}]{szeifert10}
{Szeifert}, T., {Hubrig}, S., {Sch{\"o}ller}, M., {et~al.} 2010, \aap, 509, L7

\bibitem[{{Timmer} \& {Koenig}(1995)}]{timmer95}
{Timmer}, J. \& {Koenig}, M. 1995, \aap, 300, 707

\bibitem[{{Turner} {et~al.}(2001){Turner}, {Abbey}, {Arnaud}, {Balasini},
  {Barbera}, {Belsole}, {Bennie}, {Bernard}, {Bignami}, {Boer}, {Briel},
  {Butler}, {Cara}, {Chabaud}, {Cole}, {Collura}, {Conte}, {Cros}, {Denby},
  {Dhez}, {Di Coco}, {Dowson}, {Ferrando}, {Ghizzardi}, {Gianotti}, {Goodall},
  {Gretton}, {Griffiths}, {Hainaut}, {Hochedez}, {Holland}, {Jourdain},
  {Kendziorra}, {Lagostina}, {Laine}, {La Palombara}, {Lortholary}, {Lumb},
  {Marty}, {Molendi}, {Pigot}, {Poindron}, {Pounds}, {Reeves}, {Reppin},
  {Rothenflug}, {Salvetat}, {Sauvageot}, {Schmitt}, {Sembay}, {Short},
  {Spragg}, {Stephen}, {Str{\"u}der}, {Tiengo}, {Trifoglio}, {Tr{\"u}mper},
  {Vercellone}, {Vigroux}, {Villa}, {Ward}, {Whitehead}, \& {Zonca}}]{turner01}
{Turner}, M.~J.~L., {Abbey}, A., {Arnaud}, M., {et~al.} 2001, \aap, 365, L27

\bibitem[{{Uttley} {et~al.}(2002){Uttley}, {McHardy}, \&
  {Papadakis}}]{uttley02}
{Uttley}, P., {McHardy}, I.~M., \& {Papadakis}, I.~E. 2002, \mnras, 332, 231

\bibitem[{{Vuong} {et~al.}(2003){Vuong}, {Montmerle}, {Grosso}, {Feigelson},
  {Verstraete}, \& {Ozawa}}]{vuong03}
{Vuong}, M.~H., {Montmerle}, T., {Grosso}, N., {et~al.} 2003, \aap, 408, 581

\end{thebibliography}

\phantomsection
\addcontentsline{toc}{chapter}{\appendixname}
\begin{appendix}

\section{Correcting OM fast-mode light curve of spurious features
caused by spacecraft drifts and missing OM tracking history file}
\label{appendix:drifts}

Before each OM science observation, a short-exposure image,
called a Field AcQuisition (FAQ), is taken with the $V$-filter to
identify proper field stars by comparison with the uploaded guide-star catalog.
These good guide stars are then used as a reference frame to center the
fast-mode window (composed of $22\times23$ pixels, each pixel
having a size of $\sim0\farcs48$, i.e.,
covering a field-of-view of about $10\farcs5\times11\arcsec$) 
on the target at the start of the science exposure, and to monitor any
spacecraft drifts during the OM science observation. However, owing to telemetry
constraints, no shift-and-adds of the fast-mode frames are performed on
the spacecraft. Corrections of spacecraft drifts are made on the
ground using the OM tracking history file ({\tt *OM*THX.FIT}). When no
good guide stars are found during the FAQ, the target cannot be
centered on the fast-mode window, and any spacecraft drift during the
exposure moves the target on the detector, which produces flux losses
when aperture photometry is performed.

During our observation of \ex~with the OM fast mode, no good guide
stars were found during the FAQ; therefore, the OM tracking
history file was not written to the Observation Data Files. For
illustration purposes, we focused here on the S415 exposure, although several
other exposures were found to exhibit similar artifacts. 

\begin{figure}[!t]
\centering
\includegraphics[width=\columnwidth,trim= 0 0 0 20,clip=true]{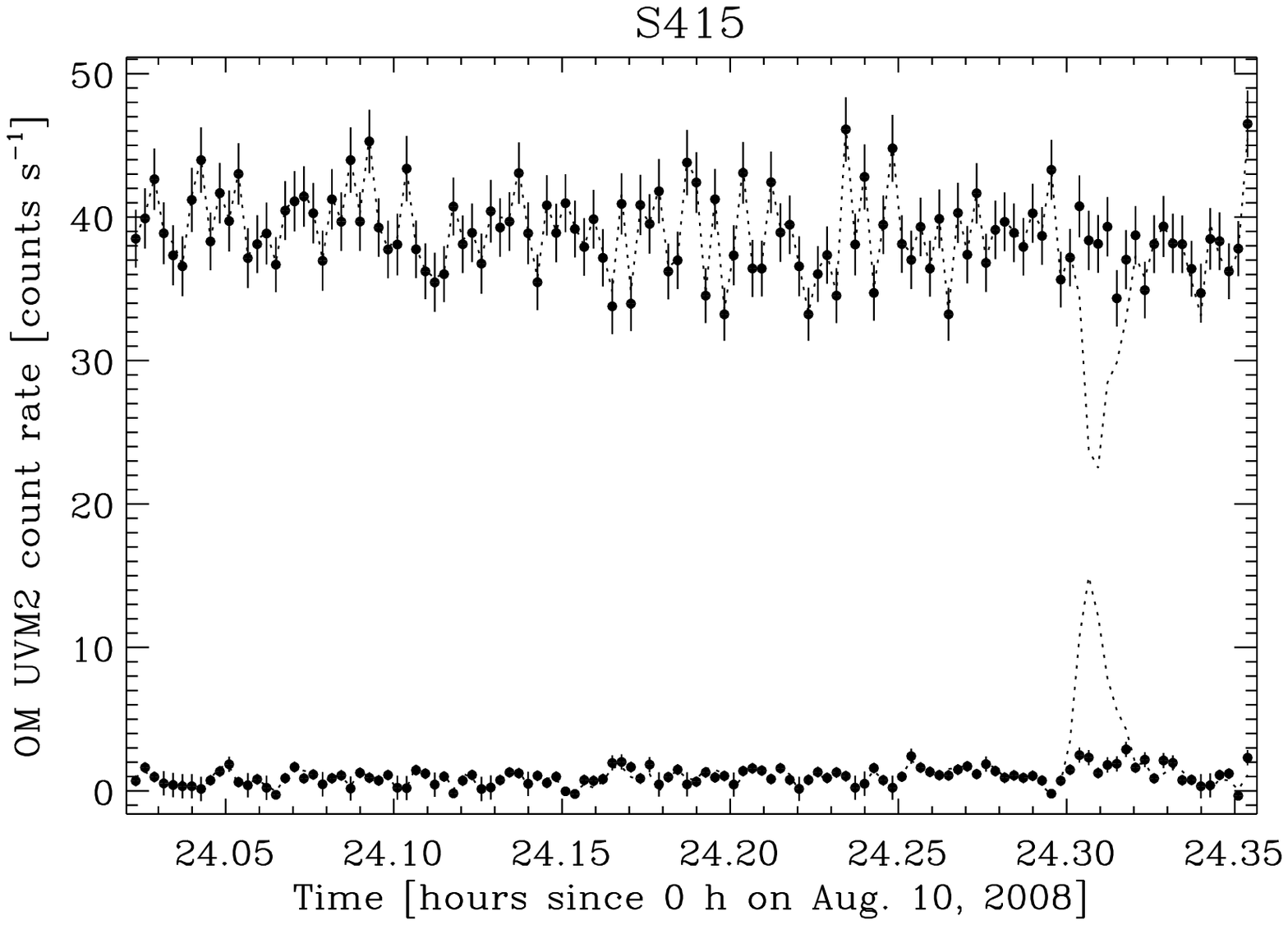}
\caption{UV light curves for \ex~(top, with the background
subtracted) and for the background (bottom) during the S415 exposure
of the OM in the fast mode. The dotted lines show the light
curves obtained with the SAS script {\tt omfchain} and no tracking
history file. The data points show the corrected light curves using our
spacecraft drift estimates. The interval of the time bin is 10~s.}
\label{figfastmode}
\vspace{0.25cm}
\centering
\includegraphics[height=\columnwidth,angle=90,trim=0 0 20 0,clip=true]{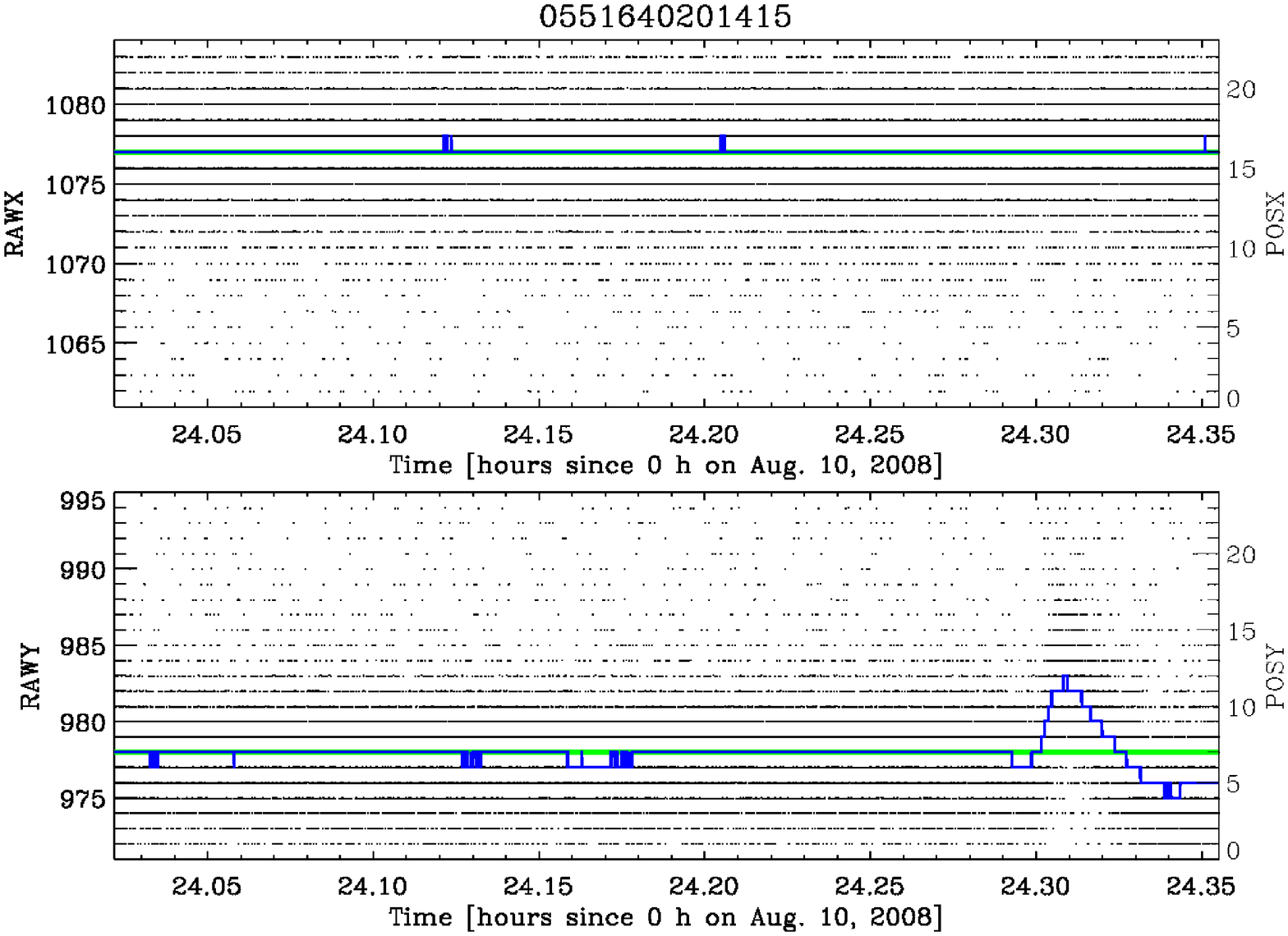}
\caption{OM fast-mode window positions versus arrival times of UV photons
detected during the S415 exposure. The top and bottom panel are the
scattered plot of the positions of the UV photons detected by the
optical monitor on the x and y-axis of the fast-mode window,
respectively, versus the frame times corresponding to the photon
arrival times. The left and right axis give the detector and the
fast window coordinates, respectively. The green line show the median
position during the exposure. The blue line shows the moving median
position versus time computed on a set of frames containing at least 500 photons. 
[\sl See the electronic edition of A\&A for a color version of this figure.]}
\label{fig:drifts}
\end{figure}

Figure~\ref{figfastmode} shows the light curves for \ex~and the
background, obtained using a 6-pixel extraction radius around \ex~and
an extraction radius of 1.2 to 2.5 times the source extraction radius
for the background.  These light curves were obtained with the SAS
script {\tt omfchain}, which assumes here zero spacecraft drifts (see
dotted lines in Fig.~\ref{figfastmode}). 
The light curves for \ex~and the background exhibit a dip and a peak,
respectively, at about 24.31~h, lasting about 1 minute, and have
identical profiles. Therefore, these features are not real, but
artifacts that can be explained by a decentering of the target with
respect to the apertures, i.e., a motion of the target on the detector
caused by spacecraft drifts.

To estimate the spacecraft drifts, we propose using the OM fast-mode
event list file ({\tt *OM*FAE.FIT}) that provides for each event its
frame number and its detector position ({\tt RAWX} and {\tt RAWY}
coordinates). We track any motion of the target on the detector versus
time by calculating the difference between the moving median
and the median value of both detector coordinates versus times, with our {\tt
IDL} program. For the moving median, we use a
set of frames containing at least 500 photons, which corresponds for
\ex~to a centered time interval with a half-width of about 6~s. The top
and bottom panels of Fig.~\ref{fig:drifts} show the scattered plots
of {\tt RAWX} and {\tt RAWY}, respectively, versus time. The green and
blue lines are the median position during the exposure and the moving
median position versus time, respectively. A drift of 5~pixels along
the detector y-axis of the target (corresponding to a motion of
$\sim2\farcs5$ on the sky), which occurred simultaneously with the features of
Fig.~\ref{figfastmode}, is visible.

To correct the OM fast-mode event list of these spacecraft drifts, we
wrote a bash script, using SAS and {\tt ftools} tasks, based
on the SAS thread entitled {\sl ``OM data reduction with SAS: step-by-step fast
mode data processing chain''\,}\footnote{This SAS thread is available at~:
\color{blue}{\tt
http://xmm.esac.esa.int/\-sas/current/documentation/threads/omf\_stepbystep.shtml}\,.}.
This SAS threads lists the 14 steps that are needed to process the fast-mode
data with SAS tasks, and describes intermediate output files.
In particular, the SAS task {\tt omfastshift} (step~6)
adds to the event list the new columns {\tt CORR\_X} and {\tt CORR\_Y}, but
using here zero drifts (i.e., {\tt CORR\_X=RAWX} and {\tt
CORR\_Y=RAWY}). Therefore, immediately after this step, we introduce
{\tt ftools} commands to add to these columns our estimate of the
spacecraft drifts. From these corrected positions, we produce a
corrected image in the detector frame from which the source detection
is performed (step~8). Figure~\ref{fig:drifts}
shows OM fast-mode light curves obtained with this method. The
spurious features due to spacecraft drifts have been corrected.

We note that the SAS task {\tt omlcbuild} (step 13) restores the
fraction of point spread function (PSF), which falls out of the
fast-mode window on the final corrected image, and not on the
individual frames. Therefore, for frames where the spacecraft drifts
are so large that the target is close to the window edges, the
fraction of PSF which falls out of the window cannot be totally
restored. For these frames, the spurious features will be only
mitigated. This limitation exists even when the OM tracking history
file is not missing.

\end{appendix}

\end{document}